\documentclass[review]{elsarticle}

\usepackage{lineno,hyperref}
\modulolinenumbers[5]

\journal{Journal of Theoretical Biology}

\usepackage{amsmath}
\usepackage{amssymb}
\newtheorem{Theorem}{Theorem}
\newproof{Proof}{Proof}

\begin{document}

\begin{frontmatter}

\title{Dynamical properties of feedback signalling in B lymphopoiesis: A mathematical modelling approach}

\author[a,b]{Salvador Chuli\'an\tnoteref{Note}}
\author[a,b]{\'Alvaro Mart\'{\i}nez-Rubio\tnoteref{Note}}
\author[c]{Anna Marciniak-Czochra}
\author[c]{Thomas Stiehl}
\author[d]{Cristina Bl\'azquez Go\~ni}
\author[d]{Juan Francisco Rodr\'iguez Guti\'errez}
\author[e]{Manuel Ramirez Orellana}
\author[e]{Ana Castillo Robleda}
\author[f,g,h]{V\'{\i}ctor M. P\'erez-Garc\'{\i}a}
\author[a,b]{Mar\'{\i}a Rosa}
\tnotetext[Note]{These authors contributed equally to this work.}

\address[a]{Department of Mathematics, Universidad de C\'{a}diz, Puerto Real, C\'{a}diz, Spain}
\address[b]{Biomedical Research and Innovation Institute of C\'adiz (INiBICA), Hospital Universitario Puerta del Mar, C\'adiz, Spain}
\address[c]{ Institute of Applied Mathematics, BioQuant and Interdisciplinary Center of Scientific Computing (IWR), Heidelberg University, Heidelberg, Germany}
\address[d]{Department of Paediatric Haematology and Oncology, Hospital de Jerez C\'adiz, Spain}
\address[e]{Department of Paediatric Haematology and Oncology, Hospital Infantil Universitario Ni\~{n}o Jes\'us, Universidad Aut\'onoma de Madrid, Madrid, Spain}
\address[f]{Department of Mathematics, Mathematical Oncology Laboratory (MOLAB), Universidad de Castilla-La Mancha, Ciudad Real, Spain}
\address[g]{Instituto de Matem\'atica Aplicada a la Ciencia y la Ingenier\'{\i}a (IMACI), Universidad de Castilla-La Mancha, Ciudad Real, Spain}
\address[h]{ETSI Industriales, Universidad de Castilla-La Mancha, Ciudad Real, Spain}

\begin{abstract}
Haematopoiesis is the process of generation of blood cells. Lymphopoiesis generates lymphocytes, the cells in charge of the adaptive immune response. Disruptions of this process are associated with diseases like leukaemia, which is especially incident in children. The characteristics of self-regulation of this process make them suitable for a mathematical study.

In this paper we develop mathematical models of lymphopoiesis using currently available data. We do this by drawing inspiration from existing structured models of cell lineage development and integrating them with paediatric bone marrow data, with special focus on regulatory mechanisms. A formal analysis of the models is carried out, giving steady states and their stability conditions. We use this analysis to obtain biologically relevant regions of the parameter space and to understand the dynamical behaviour of B-cell renovation. Finally, we use numerical simulations to obtain further insight into the influence of proliferation and maturation rates on the reconstitution of the cells in the B line. We conclude that a model including feedback regulation of cell proliferation represents a biologically plausible depiction for B-cell reconstitution in bone marrow. Research into haematological disorders could benefit from a precise dynamical description of B lymphopoiesis.
\end{abstract}


\end{frontmatter}

\linenumbers

\section{Introduction}
\begin{figure}[h!]
	\centering
	\includegraphics[width=\textwidth]{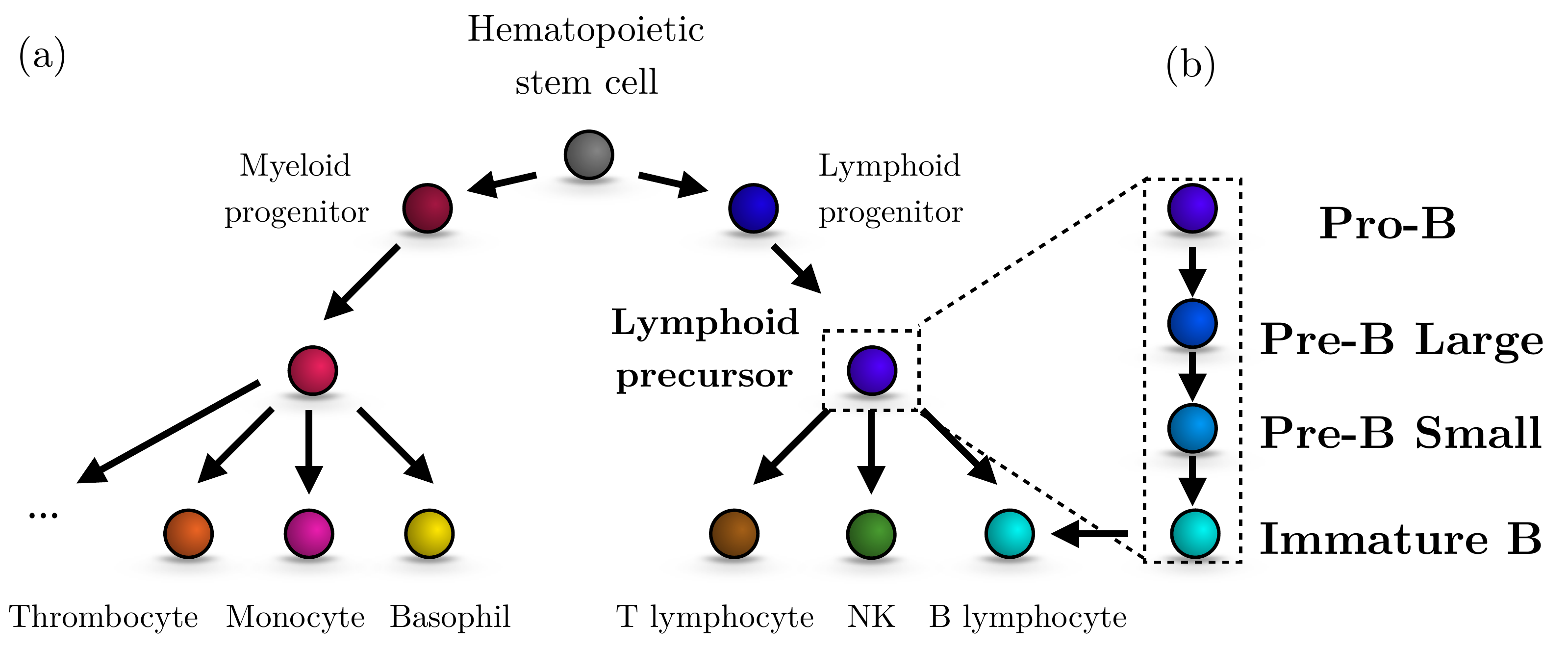}
	\caption{(a) Schematic representation of haematopoiesis. (b) B-cell lineage starts from a common lymphoid progenitor and then progresses towards immature B cells, which eventually leave the bone marrow. Hallmarks of this process are the acquisition of the immunophenotypic cell surface markers CD19 for the whole B line and CD10 for immature B cells.}
	\label{tree}
\end{figure}

Blood is a tissue under continuous regeneration, and its renovation is one of the most studied developmental processes in biology \cite{Doulatov2012}. It is initiated by haematopoietic stem cells and develops through a multi-step differentiation cascade \cite{Laurenti2018}, resulting in the generation of red blood cells, platelets and cells of the immune system. Figure \ref{tree}(a) shows the standard representation of haematopoiesis as a tree. At the top, stem cells with the potential for self-renewal give rise to respective lineage progenitors. These cells become progressively more specialised as they move towards the bottom of the tree. There are two major cell lineages: the myeloid line and the lymphoid line. The latter generates lymphocytes, involved in adaptive immune response, which is responsible for `targeted' reactions to infections.

In this work, we will focus on the description of B lymphopoiesis, i.e. the development and maturation of B cells. These cells have a range of roles, being mainly associated with the secretion of antibodies, the elements in charge of the neutralisation of foreign invaders \cite{Albert2002}. Figure \ref{tree}(b) shows a schematic representation of the route from common lymphoid progenitor to immature B cells, which eventually exit the bone marrow to complete maturation elsewhere. Alterations in the generation of B cells are related to diseases like autoimmune reactions, immunodeficiencies or lymphoproliferative disorders like lymphomas or leukaemias \cite{Lebien2008}. The latter have especial incidence in children and constitute around one third of all childhood cancer cases \cite{Steliarova2017}. 

Haematopoiesis is a perfect example of self-renewal and stemness in tissues \cite{Reya2001}. Mathematical descriptions have been performed using multi-compartmental, continuously structured models \cite{clapp2015review,Pujo2016,Lorenzi2019}. In those models, each cell type in Figure \ref{tree}(a) can be thought of as a cell compartment that receives input from the upper elements and sends its output to the lower compartments \cite{Marciniak-Czochra2009a,Roeder2006b,Viswanathan2003,Ganusov2014,Stiehl2011,Mackey1994,Dingli2009}. Some  models have focused explicitly on the B-cell line  \cite{shahaf2016b,Hu2012}. Unlike full haematopoiesis, this process is sequential, simplifying its mathematical conceptualisation. The compartmental models described above become nonlinear when the interactions between the different cell stages are included. Indeed, this process requires some kind of negative regulatory feedback in order to ensure steady production \cite{Komarova13,Manesso13}. This feature is common in the modelling of biological systems since they normally consist of a considerable number of interacting components \cite{Jones2009}. 

In the case of B cells, their development depends on the joint action of a number of factors that support or inhibit B-cell growth and differentiation \cite{Murphy16janeway}. A clear description of the participants at each stage of development is lacking \cite{Lebien2008}, which is in part due to the difficulties in recreating \textit{in-vivo} conditions in experimental designs \cite{lebien2000}. Mathematical models can help in elucidating which processes are more influenced by a given type of signal, as has been done for a number of lineages and scenarios \cite{Fuertinger13,Walenda14,Stiehl2014,Wang17,knauer2020oscillations}, and a better comprehension of these interactions can be useful in understanding progression to malignancies \cite{Jumaa2005}. 

In this context, a complete picture of B lymphopoiesis would be a useful complement to those modelling scenarios that consider an input of B cells or that represent B-cell-related phenomena \cite{Hu2012,Mehr03,Smirnova14,Mostolizadeh18,Leon20}. Precise knowledge of the dynamical behaviour of each stage of differentiation stage can also be of help in clinical situations where disorders are especially linked to the characteristics of the cell of origin. In leukaemias, for instance, the phenotype of the tumour cells is an important diagnostic criterion \cite{vandongen12}. Our aim in this paper is thus to develop a model of B lymphopoiesis in the bone marrow. We will construct and investigate the properties and dynamical behaviour of a series of models. This will be complemented with data from the literature, which mainly comes from \textit{in-vitro} assays and immune reconstitution studies, and with clinical data from haematological patients. 

This paper is structured as follows: In Sec. \ref{methods} we explain basic haematological models, going from a general model to a reduced family of models more suitable for lymphopoiesis. In Sec. \ref{maths} we perform a mathematical analysis of these models, including positivity, boundedness and stability. In Sec. \ref{results} we carry out numerical simulations, taking into account clinical data and information from the literature. In Sec. \ref{discuss} we discuss these results and examine the potential of these models to describe the process, concluding with the kind of research opportunities that this analysis paves the way for.

\section{Mathematical models and methods} 
\label{methods}

Our aim here is to describe B-cell development taking into account what the data can tell us about the structure of the population of this haematopoietic line. An illustration of the representation of cell development can be found in \cite{Marciniak-Czochra2009,stiehl2017stem},
where the authors proposed $n$ maturation stages for a cell population $u_n=u_n(t)$, with $t\in\mathbb{R}$ representing time. In this case, $u_1$ would represent stem cell population, $u_n$ a mature specialised cell and $u_i \; (i=2,\dots,n-1$) intermediate stages.

The model was studied in terms of proliferation rates $p_i=p_i(t)$ and the so-called self renewal fraction $a_i=a_i(t)$, for each maturation stage $i=1,...,n$. The latter is considered to be the probability of a cell remaining in the same cell compartment after mitosis. The authors assumed that cells at stage $i$ enter mitosis with a rate $p_i$, resulting in a total number of $2p_iu_i$ after mitosis. Then, with probability $a_i$, they remain in the same compartment, whereas with probability $1-a_i$ they go on to the next maturation stage. Therefore, each cell compartment has an output of $p_iu_i$ and an input $2p_ia_iu_i$ from their own compartment. Consequently, the previous compartment (except at the first stage) provides the input corresponding to the number of cells that go on to the next maturation stage after mitosis: $2p_{i-1}(1-a_{i-1})u_{i-1}$. Lastly, mature cells die at a rate $d$, and they are not considered to enter mitosis.  If we consider $n=3$ stages for stem cells ($u_1$), intermediate cells ($u_2$) and specialised cells ($u_3$), the result is the following system of equations:
\begin{subequations}
	\label{ModelSelfRenewal}
	\begin{align}
	\label{EarlyBEqSelfR}\dfrac{du_1}{dt}&=p_1(2a_1-1)u_1,\vspace{5pt}\\
	\label{InterBEqSelfR}\dfrac{du_2}{dt}&=p_2(2a_2-1)u_{2}+2p_{1}(1-a_{1})u_{1},\vspace{5pt}\\
	\label{MatBEqSelfR}\dfrac{du_3}{dt}&=2p_{2}(1-a_{2})u_{2}-d u_3.\vspace{5pt}
	\end{align}
\end{subequations}
B cells are far from the haematopoietic stem cells since they are already committed to the B line, thus losing part of their potential for self-renewal. We then choose to specify cell behaviour in each compartment in terms of proliferation and maturation, i.e. progression to the next stage, and to restrict the system to three different cell compartments, considering the most common immunophenotypical characterisation used in clinical practice \cite{Lochem2004}. The first compartment would also receive input from previous lymphoid progenitors. However, this early compartment is smaller and thus a constant source term contribution would be less significant \cite{Lucio1999}. Furthermore, this input is also regulated, which would require adding an equation from the previous compartment, and similarly for even earlier compartments. Our aim was to restrict the analysis to the CD19\textsuperscript{+} fraction of the B-cell line, as suggested by the data (see \ref{data}). 

Thus we will consider three compartments accounting for the different maturation stages: early B cells  ($C_1=C_1(t)$), intermediate B cells ($C_2=C_2(t)$), and finally late B cells ($C_3=C_3(t)$), where $t\in \mathbb{R}$ represents time. A compartmental model can then be written as
\begin{subequations}
	\label{Basic Model}
	\begin{align}
	\label{EarlyBEq}\dfrac{dC_1}{dt}&=\rho_1C_1-\alpha_1C_1,\vspace{5pt}\\
	\label{InterBEq}\dfrac{dC_2}{dt}&=\rho_2C_2+\alpha_1C_1-\alpha_2C_2,\vspace{5pt}\\
	\label{MatBEq}\dfrac{dC_3}{dt}&=\alpha_2C_2-\alpha_3C_3.\vspace{5pt}
	\end{align}
\end{subequations}

Note that this formulation is equivalent to the model from Eq. \eqref{ModelSelfRenewal} with $u_i=C_i$ for $i=1,2,3$ and parameters $\rho_i=p_i$, $\alpha_i=2p_i(1-a_i)$ for $i=1,2$, and $\alpha_3=d$.

Early B cells, described here by Eq. \eqref{EarlyBEq}, have a proliferation rate $\rho_1$ and a transition rate into intermediate B cells of $\alpha_1$. Analogously, the proliferation rate for intermediate B cells and transition rate into the late compartment are $\rho_2$ and $\alpha_2$, as described in Eq. \eqref{InterBEq}. In this equation, a fraction of $\alpha_1C_1$ cells comes from the early B compartment. This also happens with the $\alpha_2C_2$ cells that change their phenotypes from the intermediate B-cell into the late B-cell compartment. Late B cells in Eq. \eqref{MatBEq} are not considered to enter mitosis and they go into the blood flow with a blood transition rate $\alpha_3$.

\begin{figure}[h!]
	\centering
	\includegraphics[width=1\textwidth]{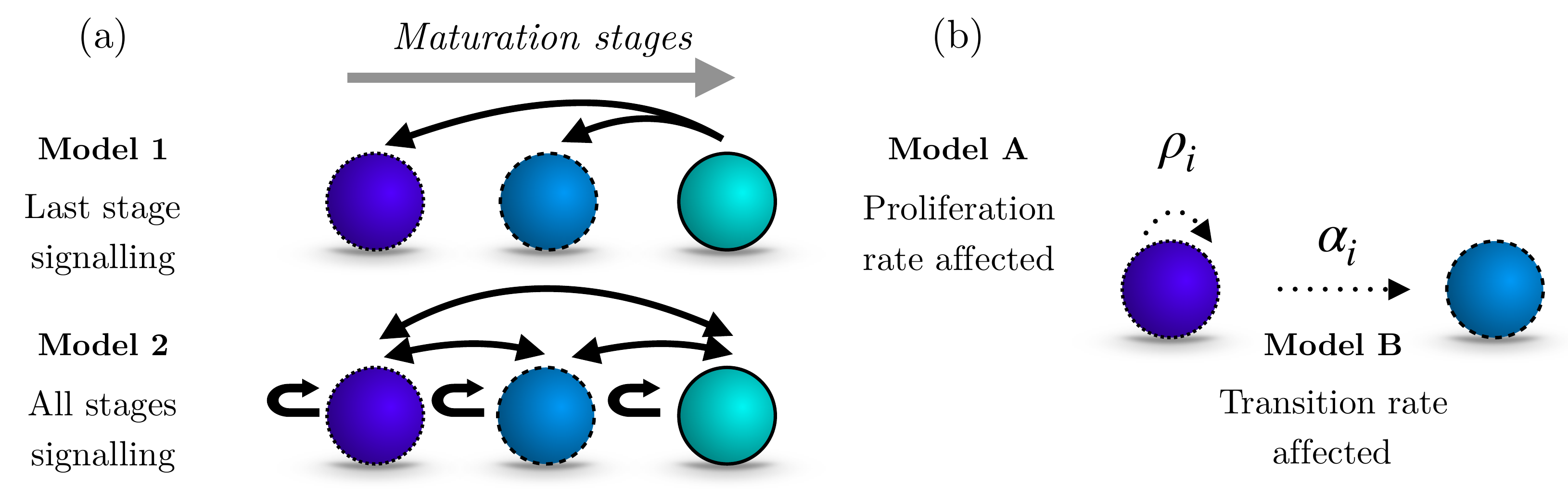}
	\caption{Representation of the different feedback signalling possibilities. (a) The signal may be produced by the most mature cells and then affect the previous compartments (model 1); or it may be produced by all cells, influencing the whole population (model 2). (b) Signalling can alternatively affect either the proliferation rate (model A) or the transition rate (model B).}
	\label{healthysignal}
\end{figure}

This compartmental model needs to be complemented with a regulatory system involving cell feedback signalling $s(t)$. Different types of signalling and the importance of the regulation of self-renewal in homeostatic (steady) state have been previously studied in the literature \cite{Stiehl2011,Komarova13,Manesso13,Fuertinger13,Walenda14,Stiehl2014,Wang17}. In our case, there are different ways of specifying which cells participate in signalling and through which processes. In this paper we consider two different hypotheses. The first is that signals are produced either by late cells (model 1) or by all cells (model 2). The second is that signals can alternatively affect either the proliferation rate (model A) or the transition rate of the model (model B). Therefore, we will consider four possibilities regarding feedback signalling. This is summarized in Figure \ref{healthysignal}.

As stated in the introduction, the precise number and role of interacting elements is unclear. The basic immune system chemical messengers are the cytokines, proteins that control cell production. They can be generated by microenvironmental elements but also from cells themselves \cite{Kraus2014}. For B cells, a number of cytokines have been shown to be relevant: IL-7 for proliferation, differentiation (transition) and survival \cite{maddaly10}, and SCF and FLT3LG \cite{oshea19} for proliferation of early stages \cite{petkau19}. Evidence in this regard comes mainly from \textit{in-vitro} and murine models and differences with humans can be significant \cite{lebien2000}. Due to this uncertainty and complexity we decided to follow an implicit formulation for the signalling as in \cite{Hu2012}, where instead of including each contributor explicitly we consider the systemic action of each, and gather them together in a single function $s(t)$.

With respect to the form of this signalling function $s(t)$, we follow the development set out in \cite{Marciniak-Czochra2009}. First, let us consider a maximal signal $\rho_S$, which is self-limited with rate $\alpha_S$. Consider then a number of cells $N=N(t)$ that regulates the production of this signal in such a way that it decreases when there is a large number. Thus, a general signalling $S=S(t)$ can be modelled as
\begin{equation}
\dfrac{dS}{dt}=\rho_S-\alpha_SS-\beta SN,
\end{equation}
where $\beta$ is the inhibitory influence of cells $N$. By making a change of variables $s(t)=\alpha_S/\rho_S\cdot S(t)$ and $k=\beta/\alpha_S$, we obtain a differential equation whose quasi-steady state is $s(t)=1/(1+kN(t))$. A rigorous analysis of this quasi-steady state approximation can be found in \cite{Marciniak18renorm}. Signal concentration depends on the number of cells of type $N$. Following the explanation in Figure \ref{healthysignal} we will consider the case when only the late cells participate in signalling, i.e.
\begin{equation}
\label{signal late}
s_1(t)=\dfrac{1}{1+k C_3},
\end{equation}
or the case where all cells participate
\begin{equation}
\label{signal all}
s_2(t)=\dfrac{1}{1+k \sum_{i=1}^3C_i}.
\end{equation}
Note that $s(t)$ is a decreasing function of the cell numbers $C_i$ so it has an inhibitory role. The parameter $k$ measures the strength of the inhibitory feedback.

To sum up, we will consider the models:
\begin{subequations}
	\label{ModelS1S2}
	\begin{align}
	\label{sapEarlyBEq}\dfrac{dC_1}{dt}&=s_\rho\rho_1C_1-s_\alpha\alpha_1C_1,\vspace{5pt}\\
	\label{sapInterBEq}\dfrac{dC_2}{dt}&=s_\rho\rho_2C_2+s_\alpha\alpha_1C_1-s_\alpha\alpha_2C_2,\vspace{5pt}\\
	\label{sapMatBEq}\dfrac{dC_3}{dt}&=s_\alpha\alpha_2C_2-s_\alpha\alpha_3C_3.\vspace{5pt}
	\end{align}
\end{subequations}
Feedback signalling affects either proliferation (signal $s_\rho=s_\rho(t)$) or transition rates (signal $s_\alpha=s_\alpha(t)$). The form of the signals $s_\rho, s_\alpha$ depends on the signalling source, either $s_1(t)$ or $s_2(t)$. This yields four possible models,  summarised in Table \ref{table1}. 
\begin{table}[!ht]
	\centering
	\begin{tabular}{lcc}
		\multicolumn{1}{c}{Signalling} & 1. Late cells & 2. All Cells \\
		\hline
		\\
		A. Affecting proliferation & Model A1 & Model A2\\
		& ($s_\rho=s_1, s_\alpha=1$) & ($s_\rho=s_2, s_\alpha=1$) \\
		\hline
		\\
		B. Affecting transition & Model B1 & Model B2 \\
		& ($s_\rho=1, s_\alpha=s_1$) & ($s_\rho=1, s_\alpha=s_2$)\\
		\hline
	\end{tabular}
	\caption{Mathematical models considered depending on signalling.}
	\label{table1}
\end{table}

\section{Theoretical results}
\label{maths}

\subsection{Existence, boundedness and positivity of solutions}

Models A can be written as
\begin{subequations}
	\label{Model A}
	\begin{align}
	\label{ProBEqProlif}\dfrac{dC_1}{dt}&=\dfrac{\rho_1C_1}{1+kN}-\alpha_1C_1,\vspace{5pt}\\
	\label{PreBEqProlif}\dfrac{dC_2}{dt}&=\dfrac{\rho_2C_2}{1+kN}+\alpha_1C_1-\alpha_2C_2,\vspace{5pt}\\
	\label{MatBEqProlif}\dfrac{dC_3}{dt}&=\alpha_2C_2-\alpha_3C_3,\vspace{5pt}
	\end{align}
\end{subequations}
and models B have the form
\begin{subequations}
	\label{Model B}
	\begin{align}
	\label{ProBEqTrans}\dfrac{dC_1}{dt}&=\rho_1C_1-\dfrac{\alpha_1C_1}{1+kN},\vspace{5pt}\\
	\label{PreBEqTrans}\dfrac{dC_2}{dt}&=\rho_2C_2+\dfrac{\alpha_1C_1}{1+kN}-\dfrac{\alpha_2C_2}{1+kN},\vspace{5pt}\\
	\label{MatBEqTrans}\dfrac{dC_3}{dt}&=\dfrac{\alpha_2C_2}{1+kN}-\dfrac{\alpha_3C_3}{1+kN}.\vspace{5pt}
	\end{align}
\end{subequations}
Feedback signalling can depend on cells from the late stage  with $N=C_3$ (models A1 and B1 with signal $s_1(t)$ as in Eq. \eqref{signal late}) or on all cells with $N=\sum_{i=1}^3C_i$ (models A2 and B2 with signal $s_2(t)$ as in Eq. \eqref{signal all}).

\begin{Theorem}
	Let us consider the following set
	\begin{equation}
	Q=\{(C_1,C_2,C_3)\in\mathbb{R}^3: C_1,C_2,C_3>0 \},
	\end{equation}
	and the initial values in $Q$
	\begin{equation}
	\label{Initial Values}
	C_1(t_0)=C_1^0,\quad
	C_2(t_0)=C_2^0,\quad
	C_3(t_0)=C_3^0,
	\end{equation}
	Then, the initial value problem for either  Eqs. \eqref{Model A} or Eqs. \eqref{Model B} has a unique local-in-time solution for each $t\in[t_0-\epsilon,t_0+\epsilon]$, for some value $\epsilon>0$.
\end{Theorem}
\begin{Proof}
	The existence of a solution for systems from Eq. \eqref{Model A} and  Eq. \eqref{Model B} is guaranteed for each $(C_1,C_2,C_3)\in Q$ by continuity of the functions 
	\begin{subequations}
		\label{frhofunctions}
		\begin{align}
		f^\rho_1=f^\rho_1(C_1,C_2,C_3)&=\dfrac{\rho_1C_1}{1+kN}-\alpha_1C_1,\vspace{5pt}\\
		f^\rho_2=f^\rho_2(C_1,C_2,C_3)&=\dfrac{\rho_2C_2}{1+kN}+\alpha_1C_1-\alpha_2C_2,\vspace{5pt}\\
		f^\rho_3=f^\rho_3(C_1,C_2,C_3)&=\alpha_2C_2-\alpha_3C_3,\vspace{5pt}
		\end{align}
	\end{subequations}
	and
	\begin{subequations}
		\label{falphafunctions}
		\begin{align}
		f^\alpha_1=f^\alpha_1(C_1,C_2,C_3)&=\rho_1C_1-\dfrac{\alpha_1C_1}{1+kN}\hspace{1pt},\vspace{5pt}\\
		f^\alpha_2=f^\alpha_2(C_1,C_2,C_3)&=\rho_2C_2+\dfrac{\alpha_1C_1}{1+kN}-\dfrac{\alpha_2C_2}{1+kN}\hspace{1pt},\vspace{5pt}\\
		f^\alpha_3=f^\alpha_3(C_1,C_2,C_3)&=\dfrac{\alpha_2C_2}{1+kN}-\dfrac{\alpha_3C_3}{1+kN}\hspace{1pt},\vspace{5pt}
		\end{align}
	\end{subequations}
	where again $N=C_3$ or $N=\sum_{i=1}^{3}C_i$.
	Boundedness of the respective partial derivatives of $f^\alpha_i$ and $f^\rho_i$ for $i=1,2,3$ proves that they satisfy the Lipschitz conditions, and therefore the solutions of systems from Eq. \eqref{Model A} and Eq. \eqref{Model B} with initial values as in Eq. \eqref{Initial Values} are unique by the Picard-Lindel\"of theorem.
\end{Proof}

Henceforth we will consider that all parameters $\rho_i,\alpha_i$ and initial conditions $C_i^0$ are positive for $i=1,2,3$.

\begin{Theorem}
	The solutions of Eqs. \eqref{Model A} and Eqs. \eqref{Model B} with $(C^0_1,C^0_2,C^0_3)\in Q$ are positive.
\end{Theorem}

\begin{Proof}
	Let us consider the functions $f^\rho_i,f^\alpha_i$ with $i=1,2,3$ from Eqs. \eqref{frhofunctions} and  Eqs. \eqref{falphafunctions}, respectively. As $f^\rho_i,f^\alpha_i>-\alpha_iC_i$, we know that
	\begin{equation}
	\dfrac{dC_i}{dt}>-\alpha_iC_i.
	\end{equation}
	By integrating both sides of the equation from $t_0$ to $t$ we obtain
	\begin{equation}
	C_i(t)>C_i^0\exp(-\alpha_it)>0.
	\end{equation}
	And therefore all solutions $C_i(t)$, $i=1,2,3$ are positive over their domain of definition. 
\end{Proof}

\begin{Theorem}
	The solutions $C_1(t), C_2(t), C_3(t)$ of Eqs. \eqref{Model A} with $(C^0_1,C^0_2,C^0_3)\in Q$ are bounded.
\end{Theorem}

\begin{Proof}
	In model from Eq. \eqref{Model A} if we consider $C_1$ to be unbounded, then
	\begin{equation}
	\lim_{C_1\rightarrow\infty} \dfrac{dC_2}{dt}=\infty,
	\end{equation}
	which would imply that $C_2$ would also be unbounded, and analogously for the case of $C_3$. We could then write
	\begin{equation}
	\label{limit}
	\lim\limits_{C_i\rightarrow\infty}\dfrac{C_i}{1+kC_3}=\lim\limits_{C_i\rightarrow\infty}\dfrac{C_i}{1+k\sum_{j=1}^3C_j}=\dfrac{1}{k}  \quad\text{ for }i=1,2,3.\\
	\end{equation}
	Let us now consider the functions $f_i^\rho$ from Eq. \eqref{frhofunctions}. From Eq. \eqref{limit} we get
	\begin{equation}
	f_1^\rho<\dfrac{\rho_1}{k}-\alpha_1C_1,\end{equation}
	and then  for all $t$
	\begin{equation}
	C_1<A e^{-\alpha_1 t}+\frac{\rho_1}{k \alpha_1},\quad A\in\mathbb{R},
	\end{equation}
	which yields that $C_1(t)$ is bounded. Considering $C_1(t)<M_1\in\mathbb{R}$, for all $t$, then 
	\begin{equation}
	f_2^\rho<\dfrac{\rho_2}{k}+\alpha_1M_1-\alpha_2C_2,
	\end{equation}
	and integrating as above implies $C_2$ is bounded.  Considering $C_2<M_2\in\mathbb{R}$, 
	\begin{equation}
	f_3^\rho<\alpha_2M_2-\alpha_3C_3,
	\end{equation}
	and thus $C_3$  is also bounded. 
\end{Proof}

For the models B, ruled by Eqs. \eqref{Model B}, we can sum the three equations to obtain 
\begin{equation}
\frac{dC_T}{dt} = \rho_1 C_1 + \rho_2 C_2 - \dfrac{\alpha_3C_3}{1+kN},
\label{boundedB}
\end{equation}
where $C_T = C_1 + C_2+C_3$. It is clear that for $C_1$ and $C_2$ sufficiently large the negative term makes only a small contribution and thus solutions are not bounded. This implies that models in which signalling affects only transition rates are not appropriate for representing biological processes of this kind.

\subsection{Steady States and stability conditions}

Biological processes in homeostasis are stable and robust. In addition to being mathematically well posed, we need to ensure that the models have a positive stable equilibrium in which the three populations coexist. In this section we study the existence of such states and their local stability. We focus on models A (Eq. \eqref{Model A}), since models B (Eq. \eqref{Model B}) do not lead to biologically relevant dynamics. An expanded analysis of the stability conditions for certain steady states of models A can be found in \ref{Model A Appendix}. Further analysis of models B, showing that they have only unstable non-trivial positive equilibria, is presented in \ref{Model B Appendix}.

\subsubsection{Model A1} 
\label{Section Model A1}
Let us consider last stage signalling $s_\rho=s_1(t)$ as in Eq. \eqref{signal late} affecting the proliferation term, i.e. we study Eq. \eqref{Model A} with $N=C_3$. The three steady states for this model are 
\begin{subequations}
	\label{steadyA1}
	\begin{align}
	P_1^{A1}&=(0,0,0),\\
	P_2^{A1}&=\left(0,\dfrac{\alpha_3(\rho_2-\alpha_2)}{k\alpha_2^2},\dfrac{\rho_2-\alpha_2}{k\alpha_2}\right),\\
	P_3^{A1}&=\left(\dfrac{\alpha_3(\rho_1-\alpha_1)(\alpha_2\rho_1-\alpha_1\rho_2)}{k\alpha_1^2\alpha_2\rho_1},\dfrac{\alpha_3(\rho_1-\alpha_1)}{k\alpha_1\alpha_2},\dfrac{\rho_1-\alpha_1}{k\alpha_1}\right).\label{SA11}
	\end{align}
\end{subequations}
The Jacobian matrix of the system at any point $(C_1,C_2,C_3)$ is
\begin{equation}
\label{JA1}
J_{A1}(C_1,C_2,C_3)=\left(
\begin{array}{ccc}
\dfrac{\rho_1}{C_3 k+1}-\alpha_1& 0 & -\dfrac{C_1 k
	\rho_1}{(C_3 k+1)^2} 
\vspace{2pt}\\
\alpha_1& \dfrac{\rho_2}{C_3 k+1}-\alpha_2 &
-\dfrac{C_2 k \rho_2}{(C_3 k+1)^2} \\
0 & \alpha_2 & -\alpha_3\\
\end{array}
\right).
\end{equation}
Substituting  $P_1^{A1} = (0,0,0)$  in Eq. \eqref{JA1} we get the eigenvalues
\begin{subequations}
	\begin{flalign}
	\lambda_{1,1}^{A1}&=-\alpha_3,\\
	\lambda_{1,2}^{A1}&=\rho_1-\alpha_1,\\
	\lambda_{1,3}^{A1}&=\rho_2-\alpha_2.
	\end{flalign}
\end{subequations}
This is the trivial equilibrium that would be unstable in normal homeostatic processes. Instability conditions are
\begin{equation}\label{cond1}
\rho_1 > \alpha_1, \text{or} \ \rho_2 > \alpha_2.
\end{equation}
As we consider $P_2^{A1}>0$, it must be $\rho_2 > \alpha_2$, and therefore $P_1^{A1}$ is unstable.

For  $P_2^{A1} $, we obtain 
\begin{subequations}
	\begin{flalign}
	\lambda_{2,1}^{A1}&=\frac{\alpha_2 \rho_1}{\rho_2}-\alpha_1,\\
	\lambda_{2,2}^{A1}&=-\frac{\alpha_3}{2}-\frac{
		\sqrt{\alpha_3 (4 \alpha_2^2-4 \alpha_2 \rho_2+\alpha_3 \rho_2)}}{2  	\sqrt{\rho_2}},\\
	\lambda_{2,3}^{A1}&=-\frac{\alpha_3}{2}+\frac{
		\sqrt{\alpha_3 (4 \alpha_2^2-4 \alpha_2 \rho_2+\alpha_3 \rho_2)}}{2  	\sqrt{\rho_2}}.
	\end{flalign}
\end{subequations}
This equilibrium point corresponds to a situation where the less differentiated compartment disappears and the system is maintained only by the proliferation of the second, leading to mature cells. As before, this is not a biologically feasible situation, thus this equilibrium must be unstable. Since $\mathcal{R}(\lambda_{2,2})<0$ and $\mathcal{R}(\lambda_{2,3})<0$, then $\frac{\alpha_2 \rho_1}{\rho_2}-\alpha_1>0$, which means that for $P_2^{A1}$ to be unstable we must have
\begin{equation}\label{cond2}
\alpha_2 \rho_1-\alpha_1\rho_2>0.
\end{equation}
From the  positivity of the non-trivial equilibrium point $P_3^{A1}$ we require
\begin{subequations}
	\label{constraintsA1}
	\begin{align}
	\rho_1&>\alpha_1,\\
	\dfrac{\rho_1}{\rho_2}&>\dfrac{\alpha_1}{\alpha_2}. \label{constraintsA1b}
	\end{align}
\end{subequations}
Conditions (\ref{constraintsA1b}) and \eqref{cond2} are identical which means that the existence of this positive equilibrium implies the instability of $P_2^{A1}$. Stability conditions for $P_3^{A1}$ are lengthy and can be found in \ref{Model A Appendix}.

\subsubsection{Model A2} 
\label{Section Model A2}
Model A2 is given by Eqs. \eqref{Model A} with $N=\sum_{i=1}^3C_i$.  The equilibria are
\begin{subequations}
	\label{steadyA2}
	\begin{align}
	P_1^{A2}&=(0,0,0),\\
	P_2^{A2}&=\left(0,\dfrac{\alpha_3(\rho_2-\alpha_2)}{k\alpha_2(\alpha_2+\alpha_3)},\dfrac{\rho_2-\alpha_2}{k(\alpha_2+\alpha_3)}\right),\\ \label{SA12}
	P_3^{A2}&=\left(\dfrac{\alpha_3(\rho_1-\alpha_1)(\alpha_2\rho_1-\alpha_1\rho_2)}{\alpha_1k\beta},\dfrac{\alpha_3\rho_1(\rho_1-\alpha_1)}{k\beta},\dfrac{\alpha_2\rho_1(\rho_1-\alpha_1)}{k\beta}\right),
	\end{align}
\end{subequations}
where
\begin{equation}
\label{beta}
\beta=\left(\alpha_2 \alpha_3 \rho_1 + \alpha_1 (\alpha_2 \rho_1 +\alpha_3 (\rho_1 - \rho_2))\right).
\end{equation}
The Jacobian matrix is 
\begin{equation}
\label{JA2}
J_{A2}=\left(
\begin{array}{ccc}
-\alpha_1 + (s^{-1}-kC_1) s^2 \rho_1& -C_1 k s^2 \rho_1& -C_1 k s^2 \rho_1\\
\alpha_1 - C_2 k s^2 \rho_2& -\alpha_2 + (s^{-1}-kC_2) s^2 \rho_2&-C_2 k s^2 \rho_2\\
0& \alpha_2& -\alpha_3
\\
\end{array}
\right).
\end{equation}
Substituting $P_1^{A2}$ in Eq. \eqref{JA2} we obtain 
\begin{subequations}
	\begin{flalign}
	\lambda_{1,1}^{A2}&=-\alpha_3,\\
	\lambda_{1,2}^{A2}&=\rho_1-\alpha_1,\\
	\lambda_{1,3}^{A2}&=\rho_2-\alpha_2.
	\end{flalign}
\end{subequations}
As before, for this equilibrium to be unstable we should have either $\rho_1>\alpha_1$ or/and $\rho_2 > \alpha_2$. From the positivity of $P_2^{A2}$, it must be the case that $\rho_2 > \alpha_2$.

For  $P_2^{A2}$, we obtain the eigenvalues
\begin{subequations}
	\label{eigenvalues A2}
	\begin{align}
	\lambda_{2,1}^{A2}&=\frac{\alpha_2 \rho_1}{\rho_2}-\alpha_1,\\
	\lambda_{2,2}^{A2}&=\frac{\alpha_2^2 \alpha_3 - 2 \alpha_2 \alpha_3 \rho_2 - \alpha_3^2 \rho_2+ h(\alpha_2,\alpha_3,\rho_2)}{2\rho_2(\alpha_2+\alpha_3)},\\
	\lambda_{2,3}^{A2}&=\frac{\alpha_2^2 \alpha_3 - 2 \alpha_2 \alpha_3 \rho_2 - \alpha_3^2 \rho_2- h(\alpha_2,\alpha_3,\rho_2)}{2\rho_2(\alpha_2+\alpha_3)}.
	\end{align}
\end{subequations} 
where $h=h(\alpha_2,\alpha_3,\rho_2)$ such that
\begin{equation}
\label{formula h}
h=\sqrt{\alpha_3}\sqrt{2 \alpha_2^2 \alpha_3 (\alpha_3 - 2 \rho_2) \rho_2 + 
	4 \alpha_2^3 (\alpha_3 - \rho_2) \rho_2 + \alpha_3^3 \rho_2^2 \
	+ \alpha_2^4 (\alpha_3 + 4 \rho_2)}.
\end{equation}

As in the previous model, the existence of $P_3^{A2}$ as a positive equilibrium influences the stability of $P_2^{A2}$.   For the positivity of $P_3^{A2}$ we have a first set of conditions
\begin{subequations}
	\label{cond 1 A2}
	\begin{align}
	\beta>0,\\
	\rho_1>\alpha_1,\\
	\alpha_2\rho_1>\alpha_1\rho_2;
	\end{align}
\end{subequations}
or a second set of conditions
\begin{subequations}
	\label{cond 2 A2}
	\begin{align}
	\beta<0,\\
	\rho_1<\alpha_1,\\
	\alpha_2\rho_1<\alpha_1\rho_2.
	\end{align}
\end{subequations}
Let us first consider that Eq. \eqref{cond 1 A2} holds. Then, in  Eq. \eqref{eigenvalues A2}, the eigenvalue $\lambda_{2,1}^{A2}$ is positive and $P_2^{A2}$ would be unstable. On the other hand, if Eq. \eqref{cond 2 A2} holds, then in  Eq. \eqref{eigenvalues A2} the eigenvalue $\lambda_{2,1}^{A2}<0$. Then $P_2^{A2}$ would be stable whenever 
\begin{equation}
|\mathcal{R}(h)| >\alpha_2^2 \alpha_3 -2 \alpha_2 \alpha_3 \rho_2 - \alpha_3^2 \rho_2
\end{equation}
for $h$ as defined in Eq. \eqref{formula h}.

In this case, for the stability of $P_2^{A2}$, the difference with model A1 is the existence of a denominator $\beta$ (Eq. \eqref{beta}). The stability conditions for  $P_3^{A2}$ are not shown here for reasons of space, but they are presented in \ref{Model A Appendix}. Stability conditions related to model B, as well as a summary of all the stability conditions depending on both models, can be found in \ref{Model B Appendix} and \ref{Summary Appendix}, respectively.
\section{Numerical results}
\label{results}

\subsection{Parameter estimation}
\label{Parameter estimation}
In the models presented above there are two parameters related to proliferation $\rho_i$ ($i=1,2$), three related to compartmental transitions $\alpha_i$ ($i=1,2,3$), and another related to the strength of signalling, the inhibition constant $k$. 

A direct measure of the proliferation rates for the specific subsets considered in this paper is lacking, but we can provide an estimation based on qualitative biological information. Normal B-cell development can be compared to data from autologous bone marrow transplantation. This type of transplantation is more likely to reproduce developmental ontogeny, especially if there is no prior or coadjuvant anti-B-cell therapy \cite{Guillaume98}. In this case B-cell progenitors can be detected in bone marrow as early as 1 month after transplantation \cite{Leitenberg94,Talmadge97}, and in blood after some delay \cite{Parrado97}. \textit{In-vitro} studies with mice show that proliferation rates are of the order of magnitude of days  \cite{Kraus2014,Deenen1990,Park1989}. Human lymphoid cultures suggest doubling times of 1 day \cite{Skipper70}. With respect to the relative values, current schemes for B-cell maturation indicate that upon CD19 acquisition there is sustained proliferation that decreases as the cell matures \cite{Kraus2014,Abbas94,Bendall14}. Late B cells already express B-cell receptor and a negative selection process occurs prior to this \cite{Murphy16janeway,Monroe03}. Based on this we can consider that the net production rate of intermediate cells is lower than early cells ($\rho_1>\rho_2$). 

In order to obtain values for transition rates we make use of steady state expressions given by (\ref{SA11}) or (\ref{SA12}). These values can be compared to the flow cytometry data of normal bone marrow (See \ref{data}). However, in our simulations we are measuring absolute cell counts while flow cytometry is only able to measure relative cell proportions. The positive non-trivial equilibrium $(C_1^*,C_2^*,C_3^*)$ given by (\ref{SA11}) or (\ref{SA12}) satisfies the relationships
\begin{subequations}
	\label{transeq}
	\begin{align}
	\label{transeq1}\dfrac{C_2^*}{C_3^*}&=\dfrac{\alpha_3}{\alpha_2},\\
	\label{transeq2}\dfrac{C_1^*}{C_2^*}&=\dfrac{\alpha_2\rho_1-\alpha_1\rho_2}{\alpha_1\rho_1}.
	\end{align}
\end{subequations}

\begin{figure}[!ht]
	\centering
	\includegraphics[width=0.6\textwidth]{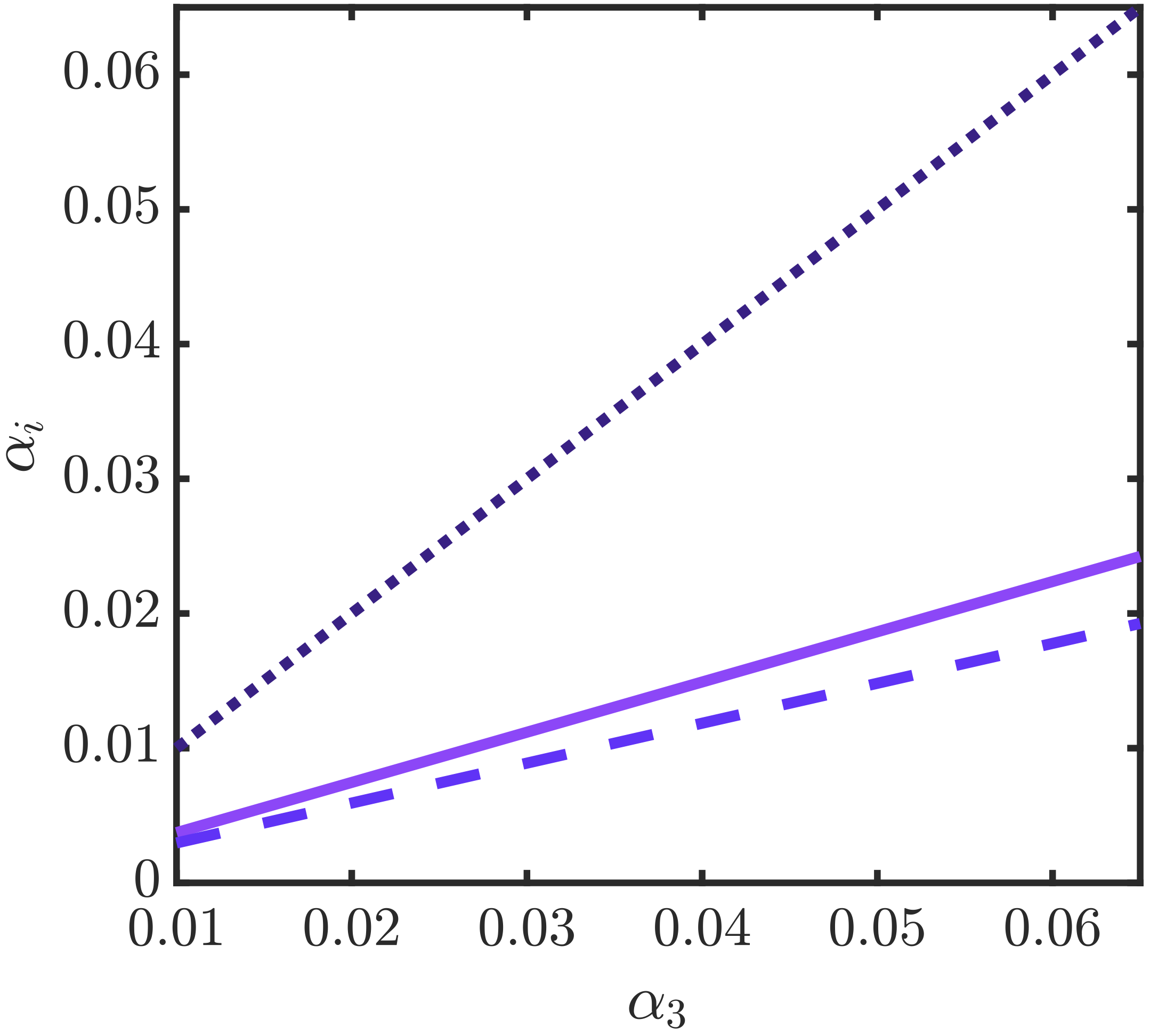}
	\caption{Dependence of the transition rates over the range of stability. Lines represent  $\alpha_1$ (light purple solid line), $\alpha_2$ (purple dashed line), $\alpha_3$ (dark purple dotted line).  Across this range $\alpha_j$ satisfies the relationship $\alpha_3>\alpha_1>\alpha_2$. Values computed from Eq. \eqref{transeq} with $\rho_1=0.0289 \; \text{h}^{-1}$, $\rho_2=0.0193 \; \text{h}^{-1}$ and steady-state cell proportions $C^*_1/C^*_2=0.0986$ and $C^*_2/C^*_3=3.227$.}
	\label{alphas}
\end{figure}

These quantities can be obtained from the analysis of the relative abundance of the three different populations. Thus, for each blood transition rate $\alpha_3$ we can obtain values of $\alpha_1,\alpha_2$ that agree with steady-state data and at the same time belong in the positive stability region of parameter space. An example of the correlation found between transition rates is shown in Figure \ref{alphas}. The order of magnitude of $\alpha_3$ can be estimated as follows. The term $\alpha_3C_3^*$ gives the total number of B cells per hour sent to blood by the bone marrow in homeostatic circumstances. In mice, bone marrow produces around $0.1\%$ of the steady state population per hour \cite{Osmond86}. In humans, the total steady-state B-lymphocyte population can be obtained from lymphocyte proportions and bone-marrow volumes from the literature and we estimate it to be $10^{10}$ cells \cite{Clark1986,Andreoni90,Caldwell91,Rego98,Nombela2017}. This yields a B-cell production of $10^7$ cells per hour, and given that $C_3^* \sim 10^9$ we estimate $\alpha_3$ to be of the order of magnitude of $10^{-2} \; \text{h}^{-1}$.

The last parameter to be determined is the inhibition constant $k$. Inspection of steady states shows that, given transition and proliferation rates of the order of magnitude of days, $k$ is of the order of magnitude of the inverse of the total steady-state population $k \propto (C_1^*+C_2^*+C_3^*)^{-1}$. The precise values for $k$ and $\alpha_3$ are selected so that we recover steady-state values and reconstitution times compatible with the literature cited above. Ranges of parameters in agreement with positivity, stability and steady-state conditions are shown in Table \ref{Table2}. 

\begin{table}[ht]
	\centering
	\begin{tabular}{|c|l|l|l|} \hline
		Param. & Meaning & Value & Units 
		\\
		\hline\hline
		
		$\rho_1$& Early B  proliferation rate&$\ln(2)$/24&hours\textsuperscript{-1}
		\\
		\hline
		$\rho_2$& Intermediate B proliferation rate&$\ln(2)$/36&hours\textsuperscript{-1}
		\\
		\hline
		$\alpha_1$& Transition rate: early to intermediate  &(0.004, 0.025)&hours\textsuperscript{-1}
		\\
		\hline
		$\alpha_2$& Transition rate: intermediate  to late  &(0.003, 0.02)&hours\textsuperscript{-1}
		\\
		\hline
		$\alpha_3$& Transition rate: late  to blood&(0.01, 0.065)&hours\textsuperscript{-1}
		\\
		\hline
		$k$&Inhibition constant &$(10^{-11}, 10^{-10})$&cells\textsuperscript{-1}
		\\ \hline
	\end{tabular}
	\caption{Parameter values for Eqs. \eqref{Model A}.}
	\label{Table2}
\end{table}

With respect to the initial state, the absolute number of mononucleated transplanted cells (MNC) is in the range of $10^9$ cells \cite{Parrado97}. From these cells, only a 1\% are B early cells ($10^7$ cells) \cite{Lucio1999,Kleiveland2015}. These cells travel through blood into the bone marrow but only 10\% of cells eventually reach the bone marrow \cite{caocci2017bone}. Therefore, we will consider for the numerical simulations of autologous transplantation an initial absolute number of cells of $C_1(0)=10^6$, $C_2(0)=0$, $C_3(0)=0$. The influence of this initial value on the dynamics of the system is described in Sec. \ref{Section Initial Value}.

\subsection{Global feedback signalling results in a smoother transition to steady states.} 

\begin{figure}[!ht]
	\centering
	\includegraphics[width=\textwidth]{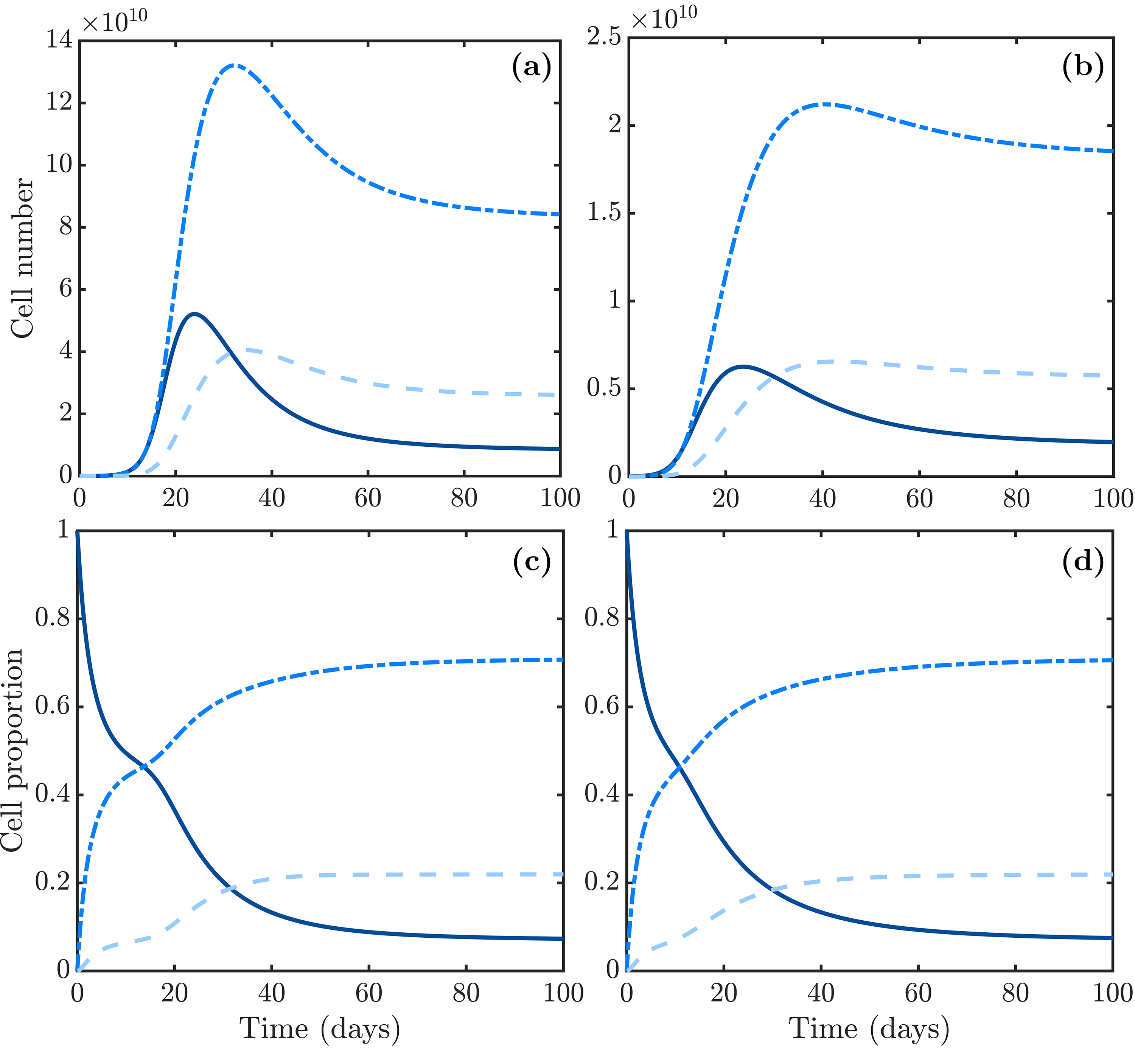}
	
	\caption{Comparison of numerical solutions of models A1 (a) and A2 (b) (Eq. \eqref{Model A}) and relative proportions (c) and (d), respectively, during the first $100$ days. Curves represent early cells $C_1$ (dark blue solid line), intermediate cells $C_2$ (blue dashed-dotted line) and late cells $C_3$ (light blue dashed line). Both simulations have initial data $C_1(0)=10^6$, $C_2(0)=0$, $C_3(0)=0$ cells and parameter values $\rho_1=0.0289 \; \text{h}^{-1}$, $\rho_2=0.0193 \; \text{h}^{-1}$, $\alpha_1=0.008 \; \text{h}^{-1}$, $\alpha_2=0.006 \; \text{h}^{-1}$, $\alpha_3=0.02 \; \text{h}^{-1}$ and $k = 10^{-10}$.}
	\label{fig all vs mat}
\end{figure}

Typical results of simulations of models A1 and A2 (Eq. \eqref{Model A}) are shown in Figure  \ref{fig all vs mat}. Recall that in model A1 (Figure  \ref{fig all vs mat}(a)) the signal depends only on cells in the final compartment while in model A2 (Figure  \ref{fig all vs mat}(b)) all cells are involved. Both models exhibit qualitatively similar behaviour. Early cells appear first, reaching a peak in population numbers slightly before day 30. Intermediate cells follow, reaching the respective peak with days of delay and with larger cell numbers. Late cells appear last and stabilize in between. The system settles into the steady state from day 80 onwards. This behaviour agrees with the conditions expressed above for the parameter values. In particular, the proportion of population from each stage is coherent with clinical data (See \ref{data}) and experimental data \cite{Lucio1999}. For both models we have 8.99\% early cells, 70.21\% intermediate cells and 20.80\% late cells.

\begin{figure}[ht]
	\centering
	\includegraphics[width=\textwidth]{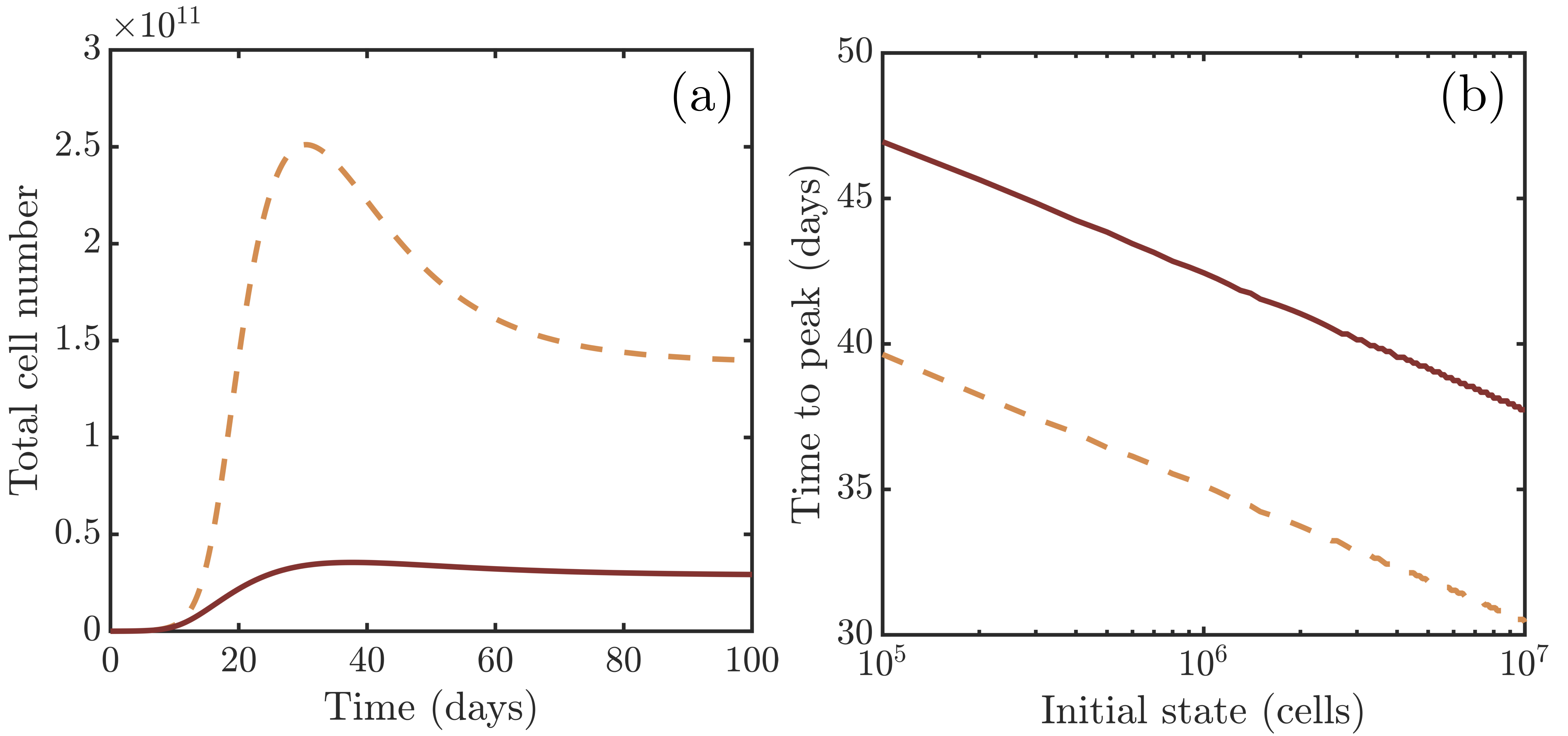}
	\caption{Comparison of numerical simulations of model A1 (orange solid line) and A2 (red dashed line) from Eq. \eqref{Model A} for: (a) total number of cells $C_1+C_2+C_3$, with initial cell numbers $C_1(0)=10^6$, $C_2(0)=0$, $C_3(0)=0$; (b) time to peak lymphocyte count as a function of initial early B cell number. Both simulations have parameter values $\rho_1=0.0289 \; \text{h}^{-1}$, $\rho_2=0.0193 \; \text{h}^{-1}$, $\alpha_1=0.008 \; \text{h}^{-1}$, $\alpha_2=0.006 \; \text{h}^{-1}$, $\alpha_3=0.02 \; \text{h}^{-1}$ and $k = 10^{-10}$.}
	\label{total}
\end{figure}

There are two main differences between the two models. The first relates to the magnitude of the steady state, which is larger when only late cells participate in signalling (model A1, Figure \ref{fig all vs mat}(a)) for the same parameter values. This can be observed in Figure \ref{total}(a), where total cell numbers for both models are shown. Note that peak lymphocyte count, i.e. the largest cell number, occurs at day 30, when intermediate cells are maximal. The second difference relates to the early behaviour of the reconstitution. In global signalling simulation (model A2, Figure \ref{fig all vs mat}(b)) there is a much less pronounced peak than when only late cells perform the signalling, presenting a smoother transition to the equilibrium state.

Figures \ref{fig all vs mat}(c) and (d) show the evolution of the percentage of each maturation stage. It is interesting to relate this to absolute counts since, as explained in Sec. \ref{Parameter estimation}, flow cytometry data only captures relative cell proportions. We observe very close behaviour between the models. The percentage of early cells quickly decreases as more mature stages appear and steady-state proportions are reached from day 80 onwards. Note that even though intermediate and mature cells have a peak in absolute cell count, this peak is not represented in terms of percentage. Also, this figure shows that a decrease in cell percentage does not necessarily means a decrease in absolute cell count, something to take into account when interpreting longitudinal flow cytometry data.

\subsection{Time to peak decreases exponentially with initial value}
\label{Section Initial Value}

In Sec. \ref{Parameter estimation} we described the rationale for the choice of the initial value of early cells. Despite this, we sought to determine how the scale of this data impacted reconstitution times. In Figure \ref{total}(b) we show the time to peak cell count for a range of initial values for early cells. There exists a decreasing exponential relationship between the two magnitudes, although the delay is not significant when considering that the literature on reconstitution after autologous transplantation describes reconstitution times in the range of 1-2 months \cite{Leitenberg94}. Multiplying initial cell numbers by 10 results in a displacement in time of 5 days.

\subsection{Blood transition rate influences time to bone marrow reconstitution}

Clinical data suggests that homeostatic bone marrow displays relatively constant subset proportions (See \ref{data}). This, together with the analysis of the expressions of the steady states allowed us to derive a connection between the three transition rates in the model (Eq. \ref{transeq}). For the ranges of parameters considered, we observed that $\alpha_3>\alpha_1>\alpha_2$ (see Figure \ref{alphas}), which suggests that the second compartment being more numerous could be due not to a higher proliferation rate but rather to a slower maturation time. This calls for the analysis of the influence of transition rates on the dynamics of the system. 

\begin{figure}[ht]
	\centering
	\includegraphics[width=\textwidth]{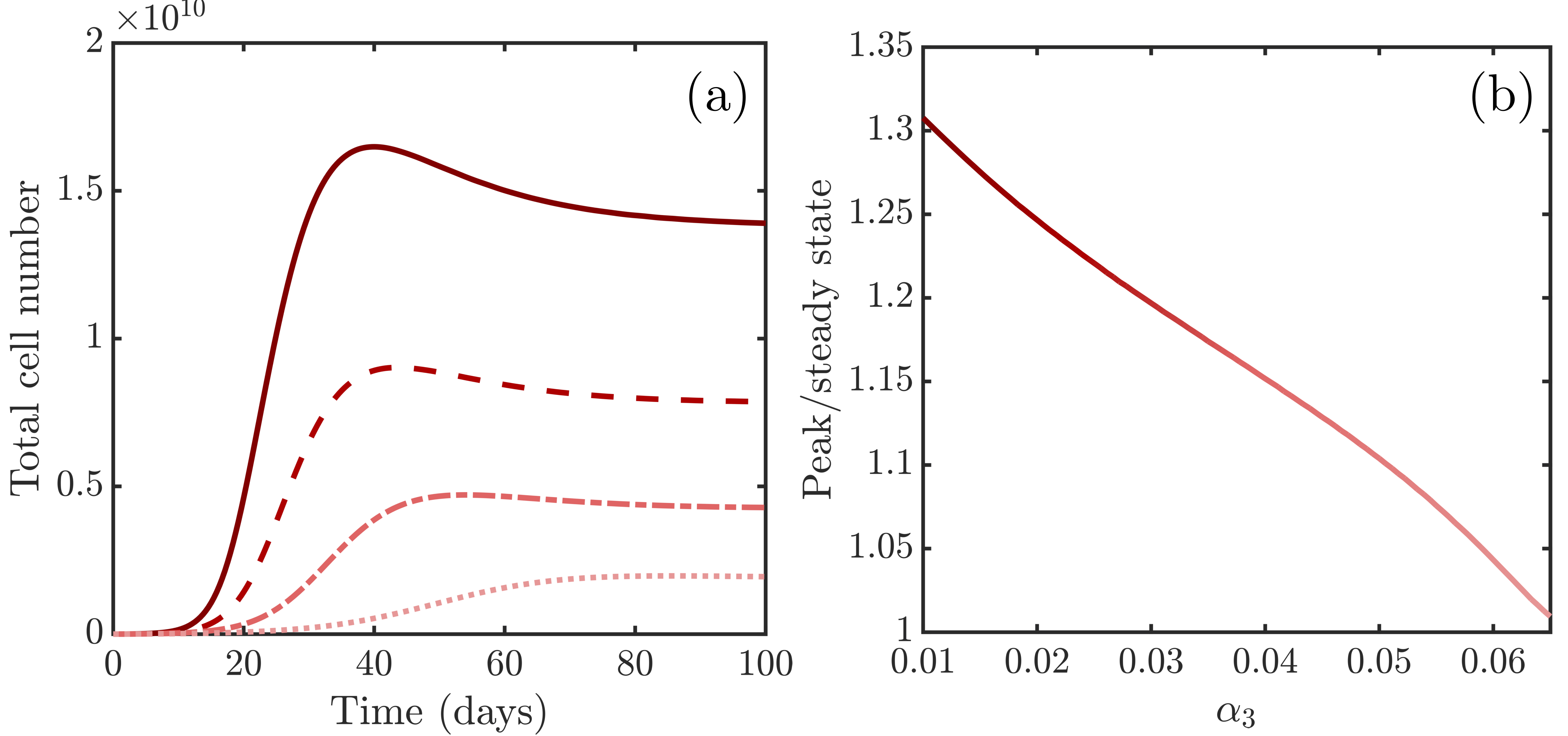}
	\caption{Numerical simulations for variable blood transition rate. (a) Total cell number $C_1+C_2+C_3$ for $t\in[0,100]$ days. Blood transition rate values: $\alpha_3=0.03 \; \text{h}^{-1}$ (solid line), $\alpha_3=0.04 \; \text{h}^{-1}$ (dashed line), $\alpha_3=0.05 \; \text{h}^{-1}$ (dashed-dotted line), $\alpha_3=0.06 \; \text{h}^{-1}$ (dotted line). Line colour goes from dark red to pink as blood transition rate increases. (b) Peak to steady state value ratio for the same range of blood transition rates. Both simulations belong to model A2 (Eq. \eqref{Model A}) with initial state $C_1(0)=10^6$, $C_2(0)=0$, $C_3(0)=0$ cells and parameter values $\rho_1=0.0289 \; \text{h}^{-1}$, $\rho_2=0.0193 \; \text{h}^{-1}$ and $k = 10^{-10}$. Parameters $\alpha_2$ and $\alpha_1$ vary with $\alpha_3$ according to Eq. \eqref{transeq}.}
	\label{fig exit rate}
\end{figure}

In order to do this we focused on variations of $\alpha_3$, the rate at which late cells exit bone marrow and enter the blood flow. We select a range of variation that lies in the positive stability region and observe the qualitative differences in the immune reconstitution. Results are shown in Figure \ref{fig exit rate}(a). 

We observe that an increasing blood transition rate means that the system reaches a lower number of total cells. Indeed, Eq. \eqref{steadyA1} shows that population levels depend on transition rates. Note that the peak during reconstitution also decreases. In order to quantify this reduction, we show, in Figure \ref{fig exit rate}(b), the proportion of the height of the peak with respect to the final steady state, for the same range of values for $\alpha_3$. For lower values of blood transition rate, the transitory population of lymphocytes can be $1.3$ times the population in homeostatic conditions. This proportion decreases if cells increase the rate at which they join blood flow. Another consequence coming from these numerical simulations is that the lower the transition rates, the longer it takes for the system to stabilise. 

\subsection{Inhibition constant has no qualitative impact on the dynamical process}

\begin{figure}[ht]
	\label{k}
	\centering
	\includegraphics[width=\textwidth]{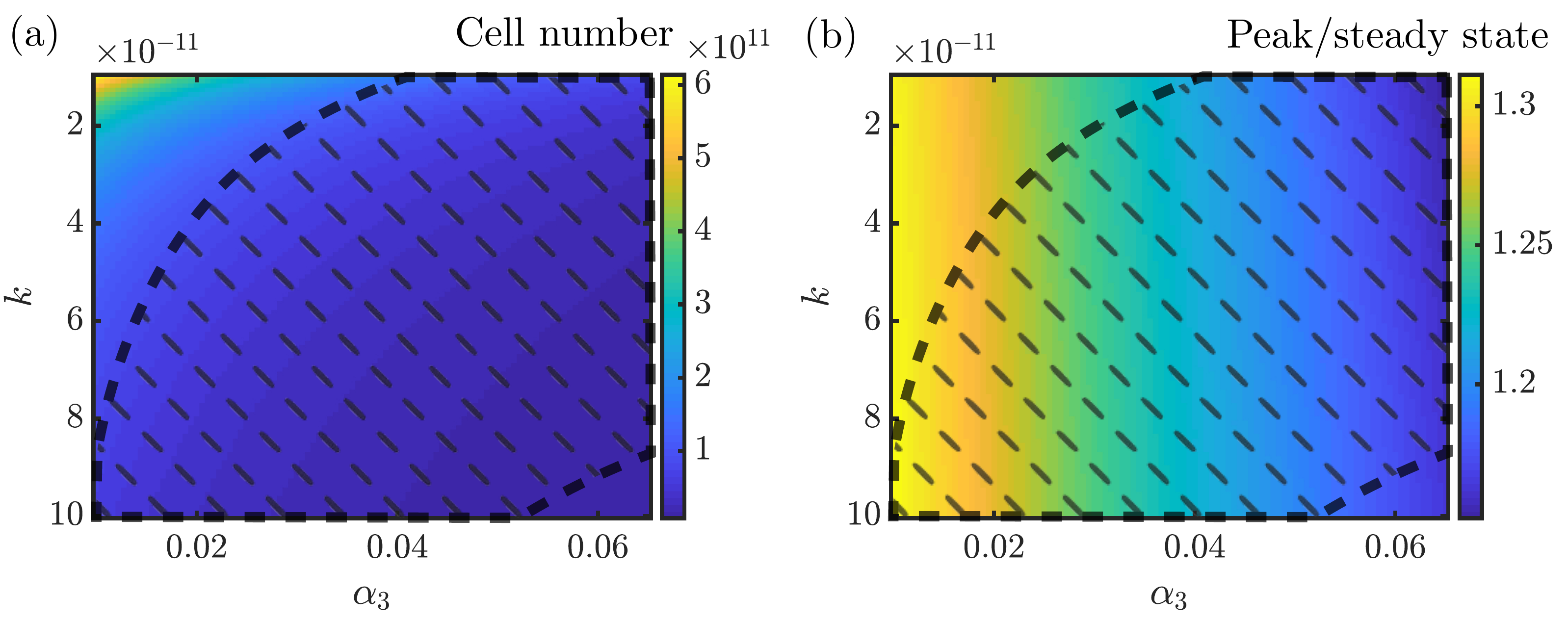}
	\caption{Influence of inhibition constant $k$ and blood transition rate $\alpha_3$ parameters in model A2 (Eq. \eqref{Model A} with all cell signalling) for (a) total steady state cell number, (b) peak vs steady states ratio. The dashed highlighted area represents the range in which total cell number $C_T\in(10^{10},10^{11})$. Each coordinate is the result of a numerical simulation of model A2 with initial state cell numbers $C_1(0)=10^6$, $C_2(0)=0$, $C_3(0)=0$ and parameter values $\rho_1=0.0289 \; \text{h}^{-1}$ and $\rho_2=0.0193 \; \text{h}^{-1}$. Parameters $\alpha_1$ and $\alpha_2$ vary with $\alpha_3\in(0.01  \text{ h}^{-1},\,0.03 \text{ h}^{-1})$ as explained in Eq. \eqref{transeq}.}
	\label{Peak}
\end{figure}

Eq. \eqref{steadyA1} shows that steady state values depend on transition rates (see also Figure \ref{fig exit rate}) but also on the inhibition constant $k$. To quantify the impact of the parameter $k$ on the dynamics of the system, we computed the total steady-state cell number as well as the proportion of peak height to steady state in a range of $\alpha_3$  (and thus $\alpha_1$ and $\alpha_2$) and $k$ values. The results are shown in Figure \ref{Peak} for model A2. With respect to the absolute final cell count, only low levels of both $k$ and $\alpha_3$ result in a much higher number of cells. We highlighted an area for which the total cell amount $C_T$ is in the range $(10^{10},10^{11})$, as estimated in Sec. \ref{Parameter estimation} from reference values. For the proportion of peak height to steady state, there is little variation in the direction of $k$, so signalling intensity has little influence on the dynamics. As shown in Figure \ref{fig exit rate}, high peak values belong in the low blood transition rate area. We conclude that transition rates primarily cause the overshoot during reconstitution, while $k$ is mainly responsible for the existence of stability regimes and the size of the final states. 

\section{Discussion and conclusions}
\label{discuss}

Haematopoiesis is one of the most widely studied biological developmental processes \cite{Doulatov2012}. Interesting questions arise related to the processes of cell lineage specification \cite{Laurenti2018,Kawamoto2010}, the role of stem cells \cite{Reya2001,Wilson2008} and the way cells communicate to regulate and ensure steady production \cite{Biasco2016,Busch2015}. This is also true for B lymphopoiesis, the branch of haematopoiesis pertaining to B-cell formation. Specific unknowns in B-cell biology are the origins of some developmental stages, the role of senescence or the array of cytokines that regulates this process \cite{Hardy07}. Studies of human B lymphopoiesis are encouraged, given that most of our knowledge about this line comes from mouse models \cite{Lebien2008}.  Answering these questions and obtaining a precise description of the dynamics of B-cell maturation are fundamental for research avenues. Some examples are the characterisation of haematological malignancies, the reconstitution after bone marrow transplant or chemotherapy or the generation of new lymphocytes from human embryonic stem cells \cite{Hardy07}.

Mathematical models have the potential to integrate biological hypotheses and clinical data in order to provide an abstract representation of biological processes \cite{Roeder2006b}. This representation can be useful not only for understanding the dynamical properties of the system, but also for testing and elucidating more inaccessible phenomena.  Our goal in this paper was to establish a biologically sensible mathematical characterization of the B lymphopoiesis. Spurred by previous models of haematopoietic processes \cite{Stiehl2011,Marciniak-Czochra2009}, we designed four models each with three differentiation stages. We added an implicit and systemic consideration of cell feedback signalling resulting in four nonlinear models.  We first analysed these models from a theoretical perspective, addressing existence, positivity, boundedness and local stability. We collected data from the literature and clinical data from haematological patients and then used numerical simulations in order to understand the role and influence of each parameter.

We learn from theoretical analysis that a stable, homeostatic situation cannot arise solely by regulating the transition rate, i.e. the process of differentiation or maturation to the next compartment. We have focused from there onward on feedback regulation of cell proliferation.  Inspection of the steady states allowed us to use flow cytometry data to establish a relationship between the three transition rates, analysing their influence by manipulating them one by one. The relationship suggests that, for the specific cell proportions in the B line, cells transition faster from early to intermediate than from intermediate to late. Intermediate cells could then be more numerous not because they proliferate more, but because of their slower transition rates downstream towards more mature cell types. Also, numerical simulations show that cell proportion is independent of which cells perform the signalling.  

In this sense, we observed that signalling coming from all stages results in a smoother reconstitution of the B cell line. Indeed, when all populations participate in signalling, their influence occurs earlier and proliferation decreases faster with time than when only late cells do. In this case we observed a peak that we understood as a consequence of the delay in the reaction of the system to overpopulation. The amplitude of this peak is also correlated with lower transition rates, which implies lower steady states values. This is biologically understandable: the late compartment saturates due to excessive input from previous compartments, delaying access to stability and thus maintaining cell production in upstream compartments. It is important to remark here that subset percentage, a common metric in follow-up samples in a clinical context, can be misleading when dynamics of expansion are at play. For example, the overshoot during early reconstitution is not observed in terms of relative proportions.  Finally, we noted that the strength of the signalling has no impact on the dynamics. The feedback loop could then be understood as a mechanism for the existence of stable output, while dynamical characteristics (time to reconstitution or early peak) are more dependent on intrinsic cellular processes. 

The idea of this study was to determine which conditions are sufficient, from a mathematical perspective, to represent the kind of biological data that is currently available for B-line development. While we obtain a behaviour that fits with the time scales of the \textit{in-vivo} process \cite{Guillaume98,Leitenberg94,Talmadge97,Parrado97,Skipper70,Lucio1999}, our study has a series of limitations. Firstly, the choice of three compartments could be refined or expanded following a more detailed characterisation of the cells. Multidimensional flow cytometry data shows that surface markers, those that specify to which stage a given cell belongs, vary continuously. A mathematical model where these markers vary continuously might be able to capture this variation. Secondly, we described signalling as a systemic phenomenon. While this was enough to recapitulate known B-cell behaviour, a more detailed description including two or more types of signalling is desirable. Lastly, the model would benefit from longitudinal data coming from immune reconstitution of the B-cell line. In this regard, flow cytometry analyses of both peripheral blood and bone marrow in routine follow-up would allow for a more precise parameterisation and enable the hypotheses presented above to be contrasted. 

To conclude, we have constructed and studied several non-linear compartmental models describing B cell lymphocyte reconstitution. These simple models describe the process of B-cell generation as portrayed by bone marrow data, and we consider it a first step in a deeper exploration of the phenomena associated with B-cell development. We verified mathematical and biological consistence, opening the door to interesting mathematical research like the existence of bifurcations or the conditions for global stability, something that finds immediate application in cases of immune reconstitution. Studies of this kind can function as a source of hypothesis generation in biomedical research, for example when contrasting mouse versus human dynamics. Ultimately, we aim to extend the methodology to situations of stability disruption and abnormal growth like B-cell disorders and other haematological diseases. 

\section*{Acknowledgements}
The support of the Junta de Andaluc\'ia group FQM-201 is gratefully acknowledged. This work has been partially supported by the Fundaci\'on Espa\~nola para la Ciencia y la Tecnolog\'ia (FECYT, project PR214 from the University of C\'adiz) and the Asociaci\'on Pablo Ugarte (APU). This work has been partially supported by the Junta de Comunidades de Castilla-La Mancha (grant number SBPLY/17/180501/000154).
\newpage
\begin{appendix}

	\section{Data}
	\label{data}
	
	\subsection{Patients}
	
	Bone marrow samples from six individuals of paediatric age (1 to 13 years) were used to estimate cell subset proportions. Four patients diagnosed with Idiopathic Thrombocytopenic Purpura (1 from Jerez Hospital and 3 from Ni\~no Jes\'us Hospital) and two patients with neutropenia (Jerez Hospital). Due to the difficulty in obtaining healthy bone marrow samples from patients of paediatric age, we selected the above as surrogate examples of normal B-cell development. Bone marrow samples were extracted from these patients in order to check for more severe disorders, but they were later diagnosed with B cell-unrelated diseases. Sample inspection further ensured lack of B-line affection. Instances of this can be found in the literature \cite{Ahmad07,Zafar18}.
	
	\subsection{Flow Cytometry}
	Bone marrow samples were analysed by flow cytometry. This technique measures the expression of the immunophenotypic markers that characterise each maturation stage. Marker expression for both hospitals' data was acquired on a FACSCanto II flow cytometer following the manufacturer's specifications Becton Dickinson (BD) for sample preparation. Samples were stained using an 8-colour panel consisting of the following six fluorochrome-conjugated antibodies provided by BD: CD38 FITC/ CD10 PE/ CD34 PerCP-Cy5-5/ CD19 PE-Cy7/ CD20 APC/ CD45 V-450. This panel allows for the identification of B cell subpopulations \cite{vandongen12}. Forward (FSC) and side scatter (SSC) were also measured.

\begin{figure}[h!]
		\centering
		\includegraphics[width=\textwidth]{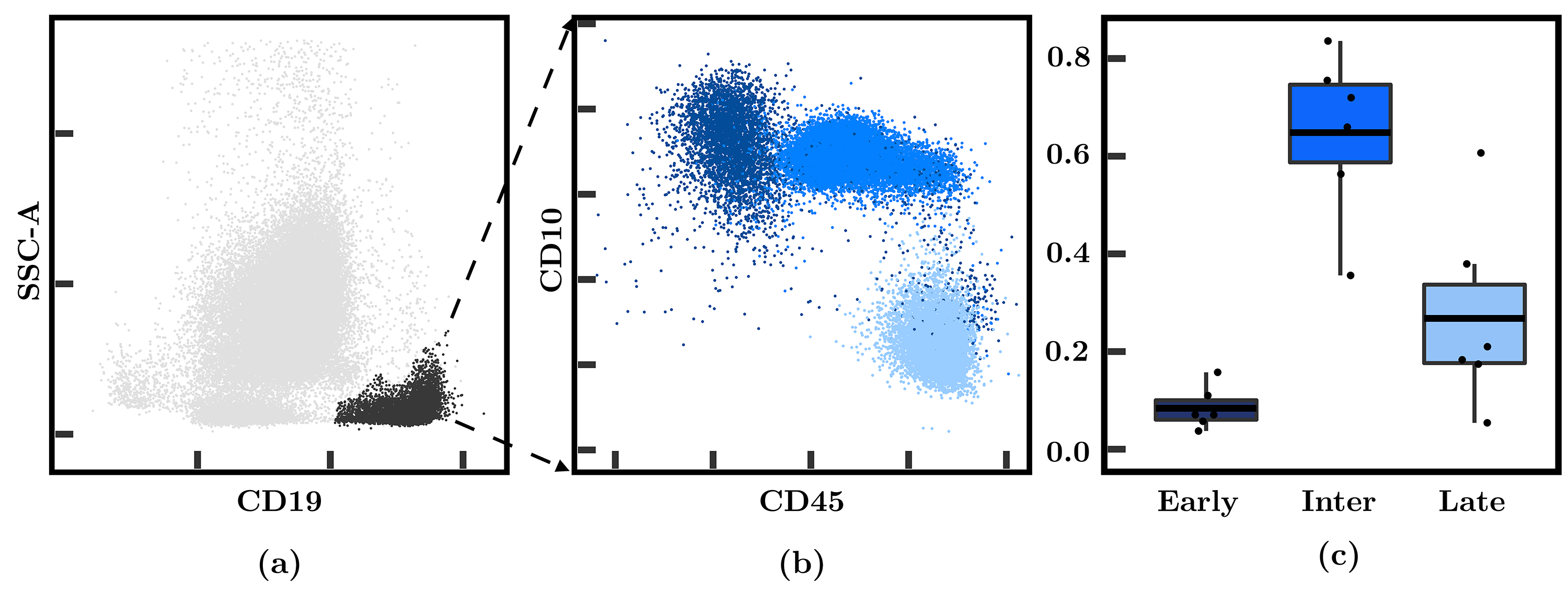}
		\caption{(a) B Lymphocyte selection in grey (CD19\textsuperscript{+} population). Y-axis represents cellular complexity, which is low for lymphocytes, and X-axis represents B Lymphocyte surface antigen CD19. (b) Within B population, three clusters were found corresponding to three maturation stages: early, intermediate and late populations. As B cells mature, they  gain expression of CD45 antigen and lose expression of CD10 antigen \cite{Lochem2004}. (c) Boxplots with proportions of each maturation stage from 6 patients displaying mean and 1st and 3rd quartile. Mean values with standard deviations: $0.083\pm 0.009, 0.648\pm 0.145, 0.268\pm 0.192$.}
		\label{fig data}
	\end{figure}

	Samples were preprocessed removing debris, doublets and marginal events as is routinely done in clinical and computational flow cytometry \cite{Saeys16}. CD19\textsuperscript{+} cells were then gated in order to select B lymphocytes \cite{Finak16}. Since the model consists of three B cell populations, we performed k-means clustering on each sample, including all B cell markers, with 3 predefined clusters. The algorithm was initialised randomly and 100 random sets were generated, selecting the one with lower within-cluster variation. Following standard immunophenotyping of B cells \cite{Lochem2004}, clusters were classified into early (CD45\textsuperscript{-}/CD10\textsuperscript{+}), intermediate (CD45\textsuperscript{+}/CD10\textsuperscript{+}) or late (CD45\textsuperscript{+}/CD10\textsuperscript{-}) B cells.  Proportions were then computed with respect to the total B lymphocyte count (CD19\textsuperscript{+} population), which correlates with experimental data \cite{Lucio1999}. Figure \ref{fig data} shows the three stages of the process. All computations were carried out in RStudio using packages flowCore \cite{Hahne2009} and flowPeaks \cite{ge2012}.
	
	\section{Stability analysis for non-trivial equilibria in model A}
	\label{Model A Appendix}
	
	\subsection{Model A1}
	We recall the model from Section \ref{Section Model A1}. From Eq. \eqref{steadyA1} we obtained the steady states $P_i^{A1}$ for $i=1,2,3$.
	
	Let us consider stability for the equilibrium point $P_3^{A2}$. We obtain the characteristic equation
	\begin{equation}
	\lambda^3+b_2\lambda^2+b_1\lambda+b_0=0,
	\end{equation}
	where
	\begin{subequations}
		\label{b_i A1}
		\begin{align}
		b_2&=\alpha_2 + \alpha_3 - \dfrac{\alpha_1 \rho_2}{\rho_1},\\
		b_1&=\alpha_3 \left(\alpha_2 - \dfrac{\alpha_1^2 \rho_2}{\rho_1^2}\right),\\
		b_0&=\dfrac{\alpha_1 \alpha_3 (\alpha_1 - \rho_1) (\alpha_1 \rho_2-\alpha_2 \rho_1)}{\rho_1^2}.
		\end{align}
	\end{subequations}
	Using the Routh-Hurwitz Criterion, for $P_3^{A1}$ to be positive and stable we must have
	\begin{equation}
	\label{Routh-Hurwitz Criterion}
	b_2b_1-b_0>0, b_2>0, b_0>0.
	\end{equation}
	The positivity conditions found in Eq. \eqref{constraintsA1} yield $b_0>0, b_2>0$. Furthermore, the stability condition $b_2b_1-b_0>0$ is satisfied if $\rho_1 \leq \rho_2$. If $\rho_1 > \rho_2$, the stability criterion is equivalent to satisfying either
	\begin{equation}
	\alpha_1^2 \rho_2 \leq \rho_1 (\alpha_1^2 - \alpha_1 \rho_1 + \alpha_2 \rho_1)
	\end{equation}
	or
	\begin{multline}
	\dfrac{\alpha_2 \rho_1^2 \big(\alpha_1^2 +  \rho_1(\alpha_2 + \alpha_3)\big) + \alpha_1^3 \rho_2^2 }{\alpha_1\rho_1}> \alpha_1 \rho_2\left(\alpha_T - \rho_1\right)  + \alpha_2 \rho_1 (\rho_1 + \rho_2),
	\end{multline}
	where $\alpha_T=\sum_{i=1}^3\alpha_i.$

	\subsection{Model A2}
	We recall the model from Section \ref{Section Model A2}. We obtained in Eq. \eqref{steadyA2} the steady states $P_i^{A2}$ for $i=1,2,3$.

	We use the Routh-Hurwitz Criterion to study the stability of $P_3^{A2}$.  We obtain the characteristic equation
	\begin{equation}
	\lambda^3+b_2\lambda^2+b_1\lambda+b_0=0,
	\end{equation}
	where
	\begin{subequations}
		\label{b_i A2}
		\begin{align}
		b_2&=\frac{(\alpha_2 \alpha_3 (\alpha_2 + \alpha_3) \rho_1^2 + \alpha_1 \
			\rho_1 (\alpha_2^2 \rho_1 + \alpha_2 \alpha_3 (3 \rho_1 - 
			2 \rho_2)}{\rho_1 \beta} +
		\\
		&\quad +\frac{\alpha_3^2 (\rho_1 - \rho_2)) - \alpha_1^2 (\
			\alpha_3 (\rho_1 - \rho_2) \rho_2 + \alpha_2 \rho_1 (\alpha_3 \
			+ \rho_2))}{\rho_1 \beta},\nonumber\\
		b_1&=\frac{\alpha_3\big(\alpha_2^2 \alpha_3 \rho_1^3 + 
			2 \alpha_1 \alpha_2 \rho_1^2 (\alpha_2 \rho_1 + \alpha_3 (\rho_1 - \rho_2)) -\alpha_1^3 \rho_1 (\rho_1 - \rho_2) (\alpha_2 + \rho_2) \big)}{\rho_1^2 \beta}+
		\\
		&\quad+\dfrac{ \alpha_3\alpha_1^4 (\rho_1 - \rho_2) \rho_2}{\rho_1^2 \beta}-\frac{\alpha_3 \alpha_1^2 \rho_1(\alpha_2^2 \rho_1 + \alpha_3 (\rho_1 - \
			\rho_2) \rho_2 + \alpha_2 \rho_1 (\alpha_3 - \rho_1 + 
			2 \rho_2))}{\rho_1^2 \beta},\nonumber\\
		b_0&=\dfrac{\alpha_1 \alpha_3 (\alpha_1 - \rho_1) (\alpha_1 \rho_2-\alpha_2 \rho_1)}{\rho_1^2}.
		\end{align}
	\end{subequations}
	
	Positivity conditions for $P_3^{A2}$ as in Eq. \eqref{cond 1 A2}  and  in Eq. \eqref{cond 2 A2} yield $b_0>0$. Finally, stability conditions $b_2>0$ and $b_2b_1-b_0>0$ result in
	\begin{subequations}
		\begin{flalign}
		&\frac{(\alpha_2 \alpha_3 (\alpha_2 + \alpha_3) \rho_1^2 + \alpha_1 \
			\rho_1 (\alpha_2^2 \rho_1 + \alpha_2 \alpha_3 (3 \rho_1 - 
			2 \rho_2)}{\rho_1 \beta} \nonumber+
		\\
		&\quad +\frac{\alpha_3^2 (\rho_1 - \rho_2)) - \alpha_1^2 (
			\alpha_3 (\rho_1 - \rho_2) \rho_2 + \alpha_2 \rho_1 (\alpha_3 \
			+ \rho_2))}{\rho_1 \beta}>0
		\end{flalign}
		and
		\begin{flalign}
		&\frac{(r_1 \rho_1^2 + r_2  \rho_1 +r_3(\rho_1 - \rho_2)  )(r_4 \rho_1^3 + r_5\rho_1^2  + r_6(\rho_1 - \rho_2) ))}{\beta^3}+\nonumber\\
		&+\alpha_1 (\rho_1-\alpha_1 ) \rho_1^2 (\alpha_1 \rho_2-\alpha_2 \rho_1 ) >0
		\end{flalign}
		where
		\begin{flalign}
		&r_1=\alpha_1 \alpha_2^2 + 3 \alpha_1 \alpha_2 \alpha_3 + \alpha_2 \alpha_3 (\alpha_2 + \alpha_3),\nonumber\\
		&r_2=-(\alpha_1^2 \alpha_2 +2 \alpha_1 \alpha_2 \alpha_3)\rho_2-\alpha_1^2 \alpha_2 \alpha_3,\nonumber\\
		&r_3=\alpha_1\alpha_3(\alpha_3 \rho_1 - \alpha_1 \rho_2),\nonumber\\
		&r_4=\alpha_1^2 \alpha_2 + 2 \alpha_1 \alpha_2^2 + \alpha_2^2 \alpha_3,\\
		&r_5=-\alpha_1^2 \alpha_2^2 - \alpha_1^2 \alpha_2 \alpha_3 - 2 \alpha_1^2 \alpha_2 \rho_2,\nonumber\\
		&r_6=2 \alpha_1 \alpha_2 \alpha_3 \rho_1^2 + \alpha_1^4 \rho_2 - \rho_1\alpha_1^2 (\alpha_1 \alpha_2 + \rho_2(\alpha_1 +\alpha_3) ).\nonumber
		\end{flalign}
	\end{subequations}

	\section{Stability analysis for models B}
	\label{Model B Appendix}
	
	\subsection{Model B1} Let us consider Eqs. \eqref{Model B} with last stage signalling  $s_\alpha=s_1(t)$ as in Eq. \eqref{signal late}, this is, $N=C_3$. The steady states for this model are
	
	\begin{subequations}
		\begin{align}
		P_1^{B1}&=(0,0,0),\\
		P_2^{B1}&=\left(0,\dfrac{\alpha_3(\alpha_2-\rho_2)}{k\alpha_2\rho_2},\dfrac{\alpha_2-\rho_2}{k\rho_2}\right),\\
		P_3^{B1}&=\left(\dfrac{\alpha_3(\alpha_1-\rho_1)(\alpha_2\rho_1-\alpha_1\rho_2)}{k\alpha_1\alpha_2\rho_1^2},\dfrac{\alpha_3(\alpha_1-\rho_1)}{k\alpha_2\rho_1},\dfrac{\alpha_1-\rho_1}{k\rho_1}\right).
		\end{align}
	\end{subequations}
	The Jacobian matrix is $J_{B1}(C_1,C_2,C_3)=J_{B1}$ such that
	\begin{equation}
	\label{JB1}
	J_{B1}=
	\left(
	\begin{array}{ccc}
	\rho_1-\dfrac{\alpha_1}{C_3 k+1} & 0 & \dfrac{\alpha_1 C_1 k}{(C_3	k+1)^2}\vspace{5pt} \\
	\dfrac{\alpha_1}{C_3 k+1} & \rho_2-\dfrac{\alpha_2}{C_3k+1} &
	\dfrac{\alpha_2 C_2 k-\alpha_1 C_1 k}{(C_3 k+1)^2}\vspace{5pt}
	\\
	0 & \dfrac{\alpha_2}{C_3 k+1} & -\dfrac{\alpha_2 C_2 k+\alpha_3}{(C_3
		k+1)^2} 
	\end{array}
	\right).
	\end{equation}
	Substituting $P_i^{B1}$ for $i=1,2,3$ in Eq. \eqref{JB1} we obtain the eigenvalues governing linear stability. First, for 
	$P_1^{B1}$, we obtain the same eigenvalues as for $P_1^{A1}$, i.e.
	\begin{subequations}
		\begin{flalign}
		\lambda_{1,1}^{B1}&=-\alpha_3,\\
		\lambda_{1,2}^{B1}&=\rho_1-\alpha_1,\\
		\lambda_{1,3}^{B1}&=\rho_2-\alpha_2.
		\end{flalign}
	\end{subequations}
	For $P_2^{B1}$ we get
	\begin{subequations}
		\begin{align}
		\lambda_{2,1}^{B1}&=\rho_1 - \dfrac{\alpha_1 \rho_2}{\alpha_2},\\
		\lambda_{2,2}^{B1}&=-\dfrac{\alpha_3 \rho_2 + 
			\sqrt{\alpha_3}\rho_2 \sqrt{4 \alpha_2 + \alpha_3 - 4 \rho_2}}{
			2 \alpha_2},\\
		\lambda_{2,3}^{B1}&=\dfrac{\alpha_3 \rho_2 -
			\sqrt{\alpha_3}\rho_2 \sqrt{4 \alpha_2 + \alpha_3 - 4 \rho_2}}{2 \alpha_2}.
		\end{align}
	\end{subequations}
	Finally, for  $P_3^{B1}$, we obtain the characteristic equation
	\begin{equation}
	\lambda^3+b_2\lambda^2+b_1\lambda+b_0=0,
	\end{equation}
	where
	\begin{subequations}
		\begin{align}
		b_2&=\dfrac{\alpha_2 \rho_1 + \alpha_3 \rho_1 - \alpha_1 \rho_2}{\alpha_1},\\
		b_1&=\dfrac{\alpha_3 \rho_1 (\alpha_2 \rho_1 + \rho_2(\rho_1-2 \alpha_1) 	)}{\alpha_1^2},\\
		b_0&=\dfrac{\alpha_3 (\alpha_1 - \rho_1) \rho_1^2 (	\alpha_1 \rho_2-\alpha_2 \rho_1)}{\alpha_1^3}.
		\end{align}
	\end{subequations}
	Considering positivity conditions for $P_2^{B1}$ and $P_3^{B1}$, we find that $\lambda_{1,i}^{B1}<0$ for $i=1,2,3$, and therefore $P_1^{B1}$ is always stable. From the positivity conditions we also get 
	\begin{subequations}
		\label{constraintsB1}
		\begin{align}
		\dfrac{\rho_1}{\rho_2}&>\dfrac{\alpha_1}{\alpha_2}.
		\end{align}
	\end{subequations}
	which implies $\lambda_{2,1}^{B1}>0$ and therefore $P_2^{B1}$ is unstable.
	
	Stability of this equilibrium $P_3^{B1}$ can be analysed by the Routh-Hurwitz Criterion from Eq. \eqref{Routh-Hurwitz Criterion}. However, given its own positivity conditions, we get $b_0<0$, implying $P_3^{B1}$ is always unstable.
	
	\subsection{Model B2} Let us now consider Eqs. \eqref{Model B} with signalling coming from all cellular compartments  $s_\alpha=s_1(t)$ as in Eq. \eqref{signal all}, this is, $N=\sum_{i=1}^3C_i$. The steady states of the model are 
	\begin{subequations}
		\begin{align}
		P_1^{B2}&=(0,0,0),\\
		P_2^{B2}&=\left(0,\dfrac{\alpha_3(\alpha_2-\rho_2)}{k(\alpha_2+\alpha_3)\rho_2},\dfrac{\alpha_2(\alpha_2-\rho_2)}{k(\alpha_2+\alpha_3)\rho_2}\right),\\
		P_3^{B2}&=\left(\dfrac{\alpha_3(\alpha_1-\rho_1)(\alpha_2\rho_1-\alpha_1\rho_2)}{\rho_1k\beta},\dfrac{\alpha_1\alpha_3(\alpha_1-\rho_1)}{k\beta},\dfrac{\alpha_1\alpha_2(\alpha_1-\rho_1)}{k\beta}\right),
		\end{align}
	\end{subequations}
	where
	\begin{equation}
	\beta=\left(\alpha_2 \alpha_3 \rho_1 + \alpha_1 (\alpha_2 \rho_1 +\alpha_3 (\rho_1 - \rho_2))\right).
	\end{equation}
	The Jacobian matrix of Eqs. \eqref{Model B} with signal $s$ given by Eq. \eqref{signal all} is $J_{B2}=J_{B2}(C_1,C_2,C_3)$ such that
	\begin{equation}
	\label{JB2}
	J_{B2}=s^2
	\left(
	\begin{array}{ccc}
	C_1 k \alpha_1 - \dfrac{\alpha_1}{s} + \dfrac{\rho_1}{s^2}
	& C_1 k \alpha_1
	& C_1 k \alpha_1\\
	\alpha_1 + kR_1\
	& kR_2 - \dfrac{\alpha_2}{s} + \dfrac{\rho_2}{s^2}
	& kR_2
	\\
	kR_3
	& \alpha_2 + kR_4
	& -\alpha_3 - kR_5 \\
	\end{array}
	\right),
	\end{equation}
	where
	\begin{subequations}
		\begin{align}
		R_1&=C_2 (\alpha_1+\alpha_2) + C_3 \alpha_1,\\
		R_2&=C_2\alpha_2-C_1 \alpha_1,\\
		R_3&=C_3 \alpha_3-C_2 \alpha_2,\\
		R_4&= C_1 \alpha_2+C_3( \alpha_2+ \alpha_3),\\
		R_5&=C_1 \alpha_3+C_2 (\alpha_2 + \alpha_3).
		\end{align}
	\end{subequations}
	Substituting $P_i^{B2}$ for $i=1,2,3$ in Eq. \eqref{JB2} we obtain the eigenvalues governing the linear stability. Specifically, for  $P_1^{B2}$, we again get
	\begin{subequations}
		\begin{flalign}
		\lambda_{1,1}^{B2}&=-\alpha_3,\\
		\lambda_{1,2}^{B2}&=\rho_1-\alpha_1,\\
		\lambda_{1,3}^{B2}&=\rho_2-\alpha_2.
		\end{flalign}
	\end{subequations}
	
	For  $P_2^{B2}$, we get the eigenvalues 
	\begin{subequations}
		\begin{align}
		\lambda_{2,1}^{B2}&=\rho_1 - \dfrac{\alpha_1 \rho_2}{\alpha_2},\\
		\lambda_{2,2}^{B2}&=-\frac{\alpha_3^2 \rho_2 +  \alpha_3 \rho_2^2+h^*(\alpha_2,\alpha_3,\rho_2)}{2\alpha_2(\alpha_2+\alpha_3)},\\
		\lambda_{2,3}^{B1}&=\frac{-\alpha_3^2 \rho_2 -  \alpha_3 \rho_2^2+ h^*(\alpha_2,\alpha_3,\rho_2)}{2\alpha_2(\alpha_2+\alpha_3)}.
		\end{align}
	\end{subequations} 
	where $h^*=h^*(\alpha_2,\alpha_3,\rho_2)$ such that
	\begin{equation}
	\label{h*}
	h^*=\sqrt{\alpha_3} \rho_2 \sqrt{	4 \alpha_2(\alpha_2^2 + \alpha_2 (2 \alpha_3 - \rho_2) +   \alpha_3 (\alpha_3 - 2 \rho_2)) + \alpha_3 (\alpha_3 - \rho_2)^2}.
	\end{equation}
	Finally, for  $P_3^{B2}$, we obtain the characteristic equation
	\begin{equation}
	\lambda^3+b_2\lambda^2+b_1\lambda+b_0=0,
	\end{equation}
	where
	\begin{subequations}
		\begin{align}
		b_2&=	\dfrac{\alpha_2 \alpha_3 \rho_1^2 (\alpha_2 + \alpha_3 + \rho_1) + 		\alpha_1 \rho_1 (\alpha_2^2 \rho_1 + \alpha_2 \alpha_3 (\rho_1 		- 2 \rho_2)}{\alpha_1 \beta} +\\
		&\quad+\dfrac{\alpha_3^2 (\rho_1 - \rho_2)) - \alpha_1^2 (\alpha_2 \rho_1 + \alpha_3 (\rho_1 - \rho_2))\rho_2}{\alpha_1 \beta},\nonumber\\
		b_1&=\dfrac{\alpha_3 \rho_1 (\alpha_2 \rho_1^2 (\alpha_3 \rho_1 + 	\alpha_2 (\alpha_3 + \rho_1)) + \alpha_1^3 (\rho_1 - \rho_2) 		\rho_2 }{	\alpha_1^2 \beta}+\\
		&\quad-\dfrac{ \alpha_1 \alpha_2 \rho_1 (\rho_1^2- 		2 \alpha_3 \rho_2 - \rho_1 \rho_2)  - \alpha_1^2 (\alpha_2 		\rho_1^2 + (\alpha_3 + \rho_1) (\rho_1 - \rho_2) \rho_2)}{\alpha_1^2 \beta},\nonumber\\
		b_0&=\dfrac{\alpha_3 (\alpha_1 - \rho_1) \rho_1^2 (	\alpha_1 \rho_2-\alpha_2 \rho_1)}{\alpha_1^3}.
		\end{align}
	\end{subequations}
	
	Every equilibrium stability is influenced by the positivity conditions of the other points. From the positivity of $P_2^{B2}$, we get that
	\begin{equation}
	\label{cond P_2 B2}
	\alpha_2>\rho_2.
	\end{equation}
	Two different scenarios arise from the positivity conditions of $P_3^{B2}$; either
	\begin{subequations}
		\label{cond 1 B2}
		\begin{align}
		\beta>0,\\
		\alpha_1>\rho_1,\\
		\alpha_2\rho_1>\alpha_1\rho_2;
		\end{align}
	\end{subequations}
	or
	\begin{subequations}
		\label{cond 2 B2}
		\begin{align}
		\beta<0,\\
		\alpha_1<\rho_1,\\
		\alpha_2\rho_1<\alpha_1\rho_2.
		\end{align}
	\end{subequations}
	Whenever Eq. \eqref{cond 1 B2} holds, equilibrium $P_1^{B2}$ is stable (mainly $\alpha_1<\rho_1$, as Eq.\eqref{cond P_2 B2} is true whenever $P_2^{B2}>0$). Moreover, equilibrium $P_2^{B2}$ would also be stable whenever Eq. \eqref{cond 2 B2} holds and also 
	\begin{equation}
	|\mathcal{R}(h^*)| <\alpha_3^2 \rho_2 +  \alpha_3 \rho_2^2
	\end{equation}
	where $h^*$ is defined as in Eq. \eqref{h*}.  However, the stability of $P_1^{B2}$ and $P_2^{B2}$ is biologically uninteresting. Focusing on the non-trivial state, with the above constraints Eq.\eqref{cond 1 B2} or Eq.\eqref{cond 2 B2} and the Routh-Hurtwitz criterion, we have $b_0<0$. Therefore, $P_3^{B2}$ is positive but always unstable. 
	
	\section{Summary of stability conditions}
	\label{Summary Appendix}
	We summarise in Table \ref{table3} the conclusions of the mathematical analysis regarding stability of the non-trivial state.
	
	\begin{table}[ht]
		\centering
		\begin{tabular}{c|cccc}
			Steady & Model A1 & Model A2 & Model B1 & Model B2\\
			State&$s_\rho=s_1,s_\alpha=1$&$s_\rho=s_2,s_\alpha=1$&$s_\rho=1,s_\alpha=s_1$&$s_\rho=1,s_\alpha=s_2$\\
			\hline
			$P_1^j$&Unstable&Unstable&Stable&Conditionally \\
			&&&&stable\\
			\hline
			$P_2^j$&Unstable&Conditionally&Unstable&Conditionally\\
			&&stable&&stable\\
			\hline
			$P_3^j$&Conditionally&Conditionally&Unstable&Unstable\\
			&Stable&Stable&&
		\end{tabular}
		\caption{Steady-state stability for every model from Eq. \eqref{ModelS1S2} under conditions of positivity of the non-trivial steady state. Index $j$ stands for the four models considering the different feedback regulations: $A1$ for cell proliferation regulation, all cell feedback; $A2$ for proliferation regulation, late cell feedback; $B1$ for transition rate regulation, late cell feedback; and $B2$ for transition rate regulation, all cell feedback.}
		\label{table3}
	\end{table}

\end{appendix}

\end{document}